\documentclass[twocolumn, amssymb, nobibnotes, aps, superscriptaddress,prb,floatfix]{revtex4-2}

\usepackage[colorlinks=true,linkcolor=blue,citecolor=blue,urlcolor=blue]{hyperref}
\usepackage{graphicx}
\usepackage{gensymb}
\usepackage[normalem]{ulem}
\usepackage{color}
\usepackage{multirow}
\usepackage{mathtools}
\usepackage{braket}
\usepackage{amsfonts}
\usepackage{natbib}
\usepackage{bm}
\usepackage{textcomp}
\usepackage{orcidlink}
\usepackage{subcaption}
\usepackage{adjustbox}
\usepackage{placeins}
\usepackage[normalem]{ulem}

\newcommand{\dagga}{{\phantom{\dagger}}}

\begin{document}

\title{Bath parameterization in multi-band cluster Dynamical Mean-Field Theory}

\author{Diego Florez-Ablan}
\affiliation{International School for Advanced Studies (SISSA), Via Bonomea 265, 34136 Trieste, Italy}

\author{Carlos Mejuto-Zaera}
\affiliation{Univ Toulouse, CNRS, Laboratoire de Physique Théorique, Toulouse, France.}

\author{Massimo Capone}
\affiliation{International School for Advanced Studies (SISSA), Via Bonomea 265, 34136 Trieste, Italy}
\affiliation{Istituto Officina dei Materiali (CNR-IOM), Via Bonomea 265, 34136 Trieste, Italy}

\begin{abstract}

Accurate and reliable algorithms to solve complex impurity problems are instrumental to a routine use of quantum embedding methods for material discovery. 
In this context, we employ an efficient selected configuration interaction impurity solver to investigate the role of bath discretization--- specifically, bath size and parameterization—in Hamiltonian-based cluster dynamical mean field theory (CDMFT) for the one- and two-orbital Hubbard models. We consider two- and four-site clusters for the single-orbital model and a two-site cluster for the two-orbital model.
Our results demonstrate that, for small bath sizes, the choice of parameterization can significantly influence the solution, highlighting the importance of systematic convergence checks. Comparing different bath parameterizations not only reveals the robustness of a given solution but can also provide insights into the nature of different solutions and potential instabilities of the paramagnetic state. 
We present an extensive analysis of the zero-temperature Mott transition of the paramagnetic half-filled single-band Hubbard model, benchmarking our findings against previous literature. We find that for the single-band model the dependence on parameterization is weak for the largest bath sizes accessible with ASCI, while a tendency towards a nematic solution can be seen when the bath size is small. Building on this, we extend our study to the multi-band regime, where we present the first systematic analysis at zero temperature for two orbitals and a two-site cluster. This setup allows us to assess the effect of nearest-neighbor dynamical correlations on the multi-orbital Mott transition.
In this case, some quantitative dependence on the parameterization is retained for the two-orbital model, for instance in the value of the critical interaction for a Mott transition.

\end{abstract}

\date{\today}

\maketitle

\section{Introduction}\label{sec:intro}

Quantum embedding techniques provide a versatile framework to study strongly correlated systems by dividing a quantum system into a fragment, that is treated exactly, and an environment, which can be described, determined and treated in different ways that define alternative embedding schemes~\cite{Sun2016,ayral_dynamical_2017,maier_quantum_2005}. When formulated in terms of a Hamiltonian representation, these methods typically substitute the environment with a discretized bath of non-interacting orbitals, which is determined self-consistently. The parameterization of this bath turns out to be a critical choice, which influences not only the computational cost and efficiency of the calculation but it can also affect the physical fidelity of the results. 

Dynamical Mean Field Theory (DMFT)~\cite{georges_dynamical_1996,Kotliar2006,Paul2019} is a prominent example of a quantum embedding method where Hamiltonian-based approaches are essential, especially for accessing low-temperature regimes. In DMFT, the lattice model is mapped on to an impurity model where the interacting fragment is coupled with a bath which is determined self-consistently requiring that the one-body Green's function of the impurity coincides with the local component of the lattice Green's function~\cite{georges_dynamical_1996}.

Several solvers have been developed to evaluate the impurity Green's function, including, mentioning only the Hamiltonian-based solvers, the Numerical Renormalization Group (NRG)~\cite{bulla_numerical_2008}, Exact Diagonalization (ED) ~\cite{caffarel_exact_1994,capone_solving_2007}, and various tensor network Ans\"{a}tze~\cite{nunez_fernandez_solving_2018,grundner_complex_2024}. However, NRG and tensor networks often rely on restrictive assumptions about the Hamiltonian form, and ED suffers from severe scalability limitations due to the exponential growth of the Hilbert space with system size.

In the context of ED, the frequency-dependent bath is explicitly discretized defining a finite number of bath levels (or sites), and the impurity problem is solved by diagonalizing the resulting finite Hamiltonian matrix containing both bath and impurity orbitals. Previous studies have examined the impact of bath discretization in single-site DMFT~\cite{Liebsch2012}, covering single-band and some multi-band cases. Yet, the role of the discretization becomes richer and more subtle in cluster extensions of DMFT (CDMFT) ~\cite{senechal_bath_2010}, where the impurities are multi-site, and the growth of the Hilbert space severely limits the number of bath orbitals that can be included in ED, making it difficult to control or assess the quality of the discretization. 

A promising alternative lies in Configuration Interaction (CI) methods, which exploit the sparsity of impurity Hamiltonians to enable efficient truncation of the Hilbert space~\cite{zgid_truncated_2012,go_adaptively_2017}. These approaches allow access to impurity problems beyond the reach of conventional ED, offering a better compromise between computational cost and accuracy . In this work, we leverage the flexibility of a CDMFT implementation combined with the Adaptive Sampling Configuration Interaction (ASCI) impurity solver to investigate the role of bath discretization in CDMFT ~\cite{mejuto-zaera_dynamical_2019,mejuto-zaera_efficient_2020,williams-young_parallel_2023}. 

Moreover, with the realistic simulation of correlated materials in mind, our study extends beyond single-band models. In particular, we consider a two-site cluster impurity in the two-band Hubbard model, the minimal setup that simultaneously captures multi-site and multi-band physics. Even this simplest model has been explored in only a limited number of studies ~\cite{crippa_local_2021,kita_spatial_2009}. 

In this work we focus on two key aspects of bath discretization: (i) the size of the bath and (ii) the parameterization of its degrees of freedom. 
In traditional ED-CDMFT implementations, the total number of orbitals in the impurity problem is typically limited to 12. Using ASCI, this number can be easily increased ~\cite{mejuto-zaera_dynamical_2019,mejuto-zaera_efficient_2020,bellomia_quasilocal_2024}, allowing us to test the influence of the bath size in different configurations. 
Moreover, in discretized Hamiltonian implementations of DMFT, it is necessary at every iteration to represent the bath in the discrete basis. This implies a fitting procedure to find the best discrete version of the bath dictated by the self-consistency by minimizing the distance between the two functions of frequency.  Especially for CDMFT and for large bath sizes, a completely unrestricted parameterization seriously challenges the optimization scheme as it is easily trapped in local minima.  This calls for different parameterizations which differ in the number and nature of the free independent parameters to use in the minimization, which may result in restrictions in the character of the solutions that the bath can represent.  If the fit cannot satisfactorily reproduce the form of Green's function that the impurity model attempts to explore, then the DMFT cycle will struggle to reach convergence, and, even if an apparently stable fix-point is found, the results will be less reliable.
Typical bath parameterization strategies consist in choosing a symmetry inspired representation for the bath~\cite{Capone2004-1d,Capone2006-competition,Civelli2005,koch_sum_2008,foley_coexistence_2019, bellomia_quasilocal_2024}, or using the freedom in the representation to make the fitting problem more adapted for the state of the art minimization algorithms~\cite{mejuto-zaera_efficient_2020}. In this work we examine in details how this choice influences the accuracy and stability of the solution.

The remainder of this work is organized as follows. We begin with a concise overview of the one- and two-band Hubbard models, along with our CDMFT+ASCI implementation used to solve them. Next, we introduce the various bath parameterization schemes employed in our study. Finally, we present and discuss our results for the one-band Hubbard model using two- and four-site cluster impurities, as well as for the two-band Hubbard model using a two-site cluster impurity.

\section{Model and Method}\label{sec:model}
\begin{figure*}
    \centering
    \subfloat[SDR (5 bath parameters)\label{fig:bath_param_SDR}]{\includegraphics[width=0.30\textwidth]{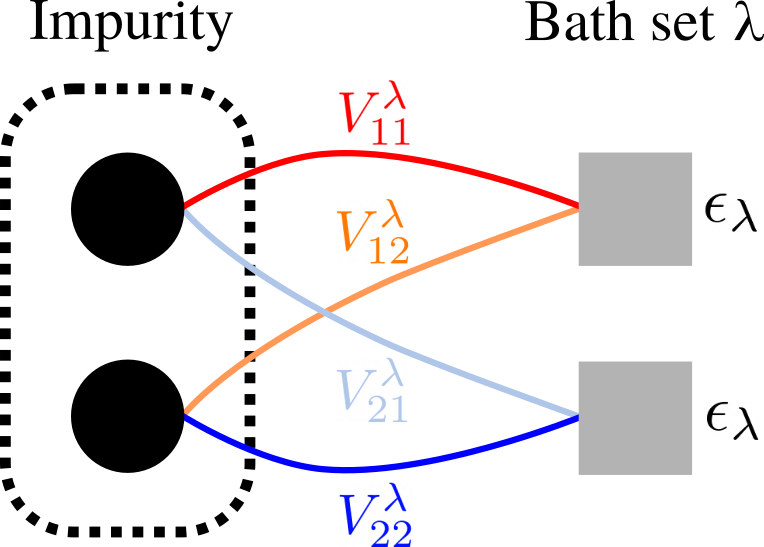}} 
    \hfill
    \subfloat[Irreps  (4 bath parameters)\label{fig:bath_param_Irrep}]{\includegraphics[width=0.30\textwidth]{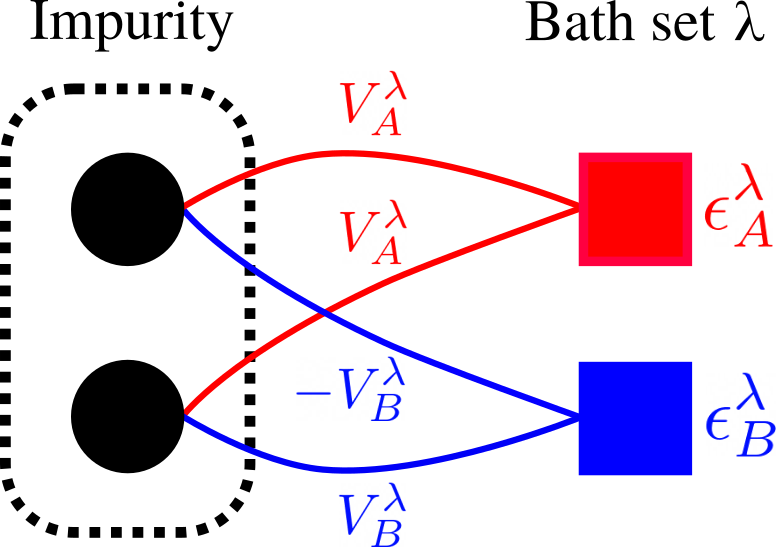}} 
    \hfill
    \subfloat[Replica  (4 bath parameters)\label{fig:bath_param_Replica}]{\includegraphics[width=0.30\textwidth]{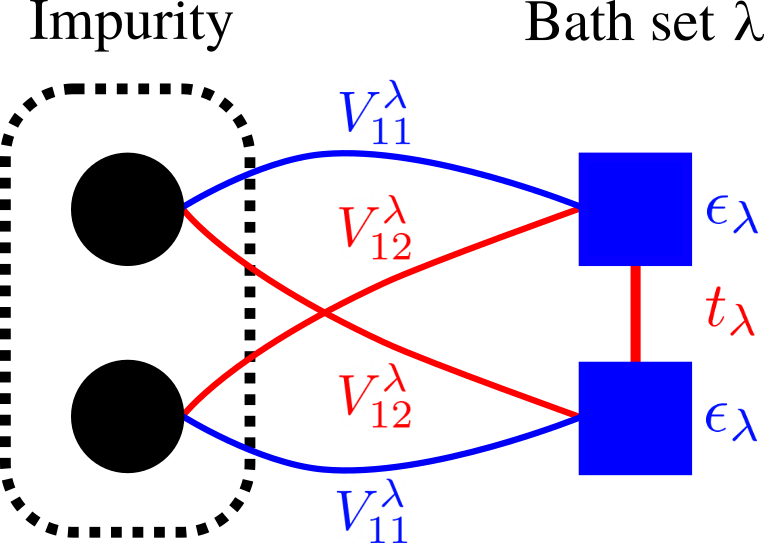}} \\
    
    \caption{ Visual representation of the different bath parameterizations in the $1\times 2$ impurity cluster for the single band Hubbard model. Here, different colors for the couplings or the bath sites indicate independent degrees of freedom. The most general bath parameterization would require $6N_{\text{sets}}$ degrees of freedom, where $N_{\text{sets}}=N_b/N_C$. }
    \label{fig:bath_params_summ}
\end{figure*}
Our reference systems are one or two-band two-dimensional Hubbard models in the square lattice with only nearest neighbor hopping at half-filling.  The single band model reads
\begin{equation}
    H^\text{1-band}=-t\sum_{\langle i,j\rangle,\sigma}\left(c_{i\sigma}^\dagger c^\dagga_{j\sigma}+H.c.\right) + U\sum_i n_{i\uparrow}n_{i\downarrow},
\end{equation}
where $c_{i\sigma}$ is the annihilation operator for the electrons, $n_{i\sigma}= c^\dagger_{i\sigma}c^\dagga_{i\sigma}$ is the number operator, $\sigma\in\{\uparrow,\downarrow\}$ iterates over the electronic spin, $U$ denotes the Coulomb repulsion, $t$ denotes the hopping amplitude and the $\langle\cdot\rangle$ brackets denote a sum over nearest-neighbors. We will adopt $t$ as the energy unit throughout this work.
On the other hand, in the 2-band case we solve the full Hubbard-Kanamori Hamiltonian, including spin-flip and pair-hopping terms:
\begin{equation}
\begin{split}
    H^\text{2-band}&=-t\sum_{\langle i,j\rangle m\sigma}c_{im\sigma}^\dagger c^\dagga_{jm\sigma}+U\sum_{im}n_{im\uparrow}n_{im\downarrow}\\&+\sum_{i\sigma\sigma'}\left(U'-\delta_{\sigma\sigma'}J\right)n_{i1\sigma}n_{i2\sigma'}\\&-J\sum_i\left(c_{i1\uparrow}^\dagger c^\dagga_{i1\downarrow} c_{i2\downarrow}^\dagger c_{i2\uparrow}^\dagga+c^\dagger_{i1\uparrow}c^\dagger_{i1\downarrow}c^\dagga_{i2\uparrow}c^\dagga_{i2\downarrow}+H.c.\right).
\end{split}
\end{equation}
With $m\in\{1,2\}$ spanning the bands, $J$ the Hund's coupling and $U'$ the inter-band Coulomb repulsion. We impose the condition $U'=U-2J$ to enforce rotational symmetry. 
The electronic dispersion for these models on the square lattice is given by $\epsilon_{\bm{k}}=-2t\left(\cos k_x + \cos k_y \right)$.
We solve these models using CDMFT in the Hamiltonian implementation. Thus, we map the lattices into impurity models containing several sites, with one or two bands, coupled to a set of non-interacting bath sites:
\begin{eqnarray}
H_\text{imp}&=&H_C+\sum_{\ell=1,\sigma}^{N_{b}}h_{\ell,\ell'}\  a^\dagger_{\ell,\sigma} a^\dagga_{\ell',\sigma} \nonumber\\&+& \sum_{\ell=1,\sigma}^{N_{b}}\sum_{\alpha=1}^{N_C}\left( V_{\alpha,\ell}\ a^\dagger_{\ell,\sigma} c^\dagga_{\alpha,\sigma} + H.c.\right)
\label{H_imp}
\end{eqnarray}
where $N_C$ is the number of cluster degrees of freedom, $N_{b}$ is the number of bath levels (or sites), $H_C$ is the original fully interacting Hamiltonian restricted to the cluster degrees of freedom, $c_{\alpha,\sigma}$ and $a_{\ell,\sigma}$ are the cluster and bath annihilation operators respectively. The index $\alpha$ collectively labels sites and orbitals.
The bath degrees of freedom are characterized by a single-particle Hamiltonian $h_{\ell,\ell'}$ and their couplings $V_{\alpha, \ell}$ to the cluster sites. The bath hybridization function is defined as 
\begin{equation}
\Delta^\text{Bath}(i\omega)_{\alpha,\beta} = 
\bm{V_{\alpha}} ({i\omega \mathbb{I}_{N_{b}}-\hat{h}})^{-1}\bm{V_{\beta}^\dagger},
\label{Delta_discrete}
\end{equation}
where $\bm{V_{\alpha}}$ is the vector of dimension $N_{b}$ that contains all the $V_{\alpha, \ell}$, and $\hat{h}$ is the matrix representation of $h_{\ell,\ell'}$, while $\mathbb{I}_{N_{b}}$ is the identity matrix in the bath space.

In this work we compute the zero-temperature Green function of this Hamiltonian using ED or ASCI. ASCI is an iterative selected CI method~\cite{Huron1973,Tubman2016,Garniron2018,tubman_modern_2020} that constructs an accurate representation of the ground-state wavefunction by selectively including the most important Slater determinants. The process begins with an initial set of determinants, which is iteratively expanded. Each determinant in the expanded set is then evaluated and ranked based on its perturbatively estimated contribution to the ground-state wavefunction. Only the most significant determinants are retained to form the basis for the next iteration. This adaptive selection and refinement process continues until the ground-state energy converges, ensuring both computational efficiency and high accuracy ~\cite{tubman_modern_2020}.
The ASCI method allows us to solve larger impurity problems at smaller computational cost, for this reason, here we use it for all calculations where the involved number of atomic orbitals is bigger than 12. Otherwise, the exact diagonalization method is used, since it remains numerically affordable.

Computing the Green function of the cluster impurity $G_\text{imp}(i\omega)_{\alpha,\beta}$ with one of these two techniques, we can then extract the self-energy of the cluster using the Dyson equation:
\begin{equation}
\begin{split}
    \Sigma_\text{imp}(i\omega)_{\alpha,\beta}&= (i\omega+\mu)\delta_{\alpha\beta}-h_{C,\alpha,\beta}\\&-G_\text{imp}^{-1}(i\omega)_{\alpha,\beta}-\Delta^\text{Bath}(i\omega)_{\alpha,\beta},
\end{split}
\end{equation}
where $h_C$ is the non-interacting part of $H_C$ and $\mu$ is the chemical potential. One can then build the lattice Green function by using the cluster self-energy as an approximation for the Green function of the lattice, this step embodies the DMFT approximation~\cite{georges_dynamical_1996}:
\begin{equation}
    G_\text{latt}(\tilde{\bm{k}},i\omega)=\left[(i\omega +\mu )-h(\tilde{\bm{k}})-\Sigma_\text{imp}(i\omega)\right]^{-1}
\end{equation}
where $h(\tilde{\bm{k}})$ is the Fourier transform of $h_C$ from the lattice to the reduced Brillouin Zone (RBZ). Integrating this equation over its momentum coordinates yields the local Green function:
\begin{equation}
    G(\bm{R}_0,i\omega)=\frac{1}{V_\text{RBZ}}\int_\text{RBZ}d\bm{k}\, G_\text{latt}(\bm{k},i\omega).
\end{equation}

From this, one can employ again the Dyson equation to the local Green function to obtain a new value for the hybridization $\Delta^\text{DMFT}(i\omega)$:
\begin{equation}
\begin{split}
    \Delta^{\text{DMFT}}(i\omega)_{\alpha,\beta}&=(i\omega+\mu)\delta_{\alpha,\beta}-h_{C,\alpha,\beta}\\
    &-G^{-1}(\bm{R}_0,i\omega)_{\alpha,\beta}-\Sigma_\text{imp}(i\omega)_{\alpha,\beta}.
\end{split}
\label{newdelta}
\end{equation}

In order to proceed with the self-consistent loop Eq.~(\ref{newdelta}) must be expressed in the discrete form (\ref{Delta_discrete}). In order to obtain the bath parameters appearing in the latter equation we perform a fitting procedure that will provide the new values of $\epsilon_\ell$ and $V_{\alpha,\ell}$ that best approximate Eq.~(\ref{newdelta}). This is the crucial step of the Hamiltonian-based solution, and this is where the bath parameterization plays a crucial role.

In practice, the fitting consists of the minimization of a distance function under some metric. In this work we define the distance as

\begin{equation}
    \mathcal{J}(\{{\epsilon}_\ell,\bm{V}_\ell\}) = \frac{1}{N_\omega}\sqrt{\sum_{n=1}^{N_\omega}\left\lvert\left\lvert {\Delta}^{\text{DMFT}}(\omega_n)-{\Delta}^{\text{Bath}}(\omega_n)\right\rvert\right\rvert^2_F}
\end{equation}

As customary, all DMFT calculations in this work were performed in the imaginary frequency axis, taking $N_\omega=10^3$ frequencies with an effective inverse temperature $\beta t=50\pi$, which sets the sampling rate for the Matsubara frequencies. After a calculation was deemed converged, we proceeded to perform real frequency calculations with a Lorentzian broadening of $\delta=0.1t$. 
In the next section we discuss in details the potential problems and subtleties of this minimization procedure and we compare different recipes. 

Most of our analysis will be based on relevant observables that can be directly computed on the cluster (cluster quantities), namely
\begin{itemize}
    \item[(i)] The quasiparticle residues $Z_{\alpha\alpha}$ from the value of the imaginary part of the self-energy at the first Matsubara frequency $i\omega_0$:
    \begin{equation}
        Z_{\alpha \alpha} = \left(1-\frac{ \text{Im} \Sigma_{\alpha\alpha} (i\omega_0)}{\omega_0}\right)^{-1}.
    \end{equation}
    
    \item[(ii)] An estimation of the spectral weight at low (real) frequency obtained by integrating the spectral function $A(\omega)$ over the interval $\omega \in \left[ -t/2,t/2\right]$.
    
    \item[(iii)] The value of the nearest-neighbour component of the real part of the self-energy at zero frequency.
    
    \item[(iv)] The double occupancies $\langle n_{i\uparrow} n_{i\downarrow} \rangle$, where $n_{i\sigma}$ is the density operator for electrons of spin $\sigma$ on site $i$.
    
    \item[(v)] Spin-spin correlation functions $\langle s^z_i s^z_j \rangle$, where $s^z_i = (n_{i\uparrow} - n_{i\downarrow})/2$.
    
    \item[(vi)] Band correlation functions (for multi-band calculations) $\langle \tau_i^z \tau_j^z \rangle$, where $\tau^z_i = \sum_\sigma \left( n_{i1\sigma} - n_{i2\sigma} \right)/2$.
\end{itemize}

Moreover, in the case of the $2\times2$ cluster impurity, where the cluster preserves the symmetry structure of the lattice, we will also present representative results for momentum-resolved lattice properties. Our primary focus is on lattice quantities that are direct outputs of our calculations and enable straightforward comparison between different solutions. Additionally, we include a section (Sec. \ref{Sec:lattice}) that presents selected results for lattice properties, in particular the momentum-resolved spectral functions, which require the choice of a periodization procedure from several available options.

\section{Bath parameterizations}

As we discussed above, the most general parameterization of the bath is Eq. (\ref{H_imp}). That corresponds to the hybridization function Eq. (\ref{Delta_discrete}). We can always diagonalize $\hat{h}$ which leaves us with a non-convex minimization problem involving $N_{b}(1+N_{C})$ real variables corresponding to the $N_{b}$ eigenvalues of $\hat{h}$, labeled $\epsilon_\ell$, and $N_{b}N_{C}$ hybridization amplitudes $V_{\alpha,\ell}$.

Being a high-dimensional, non-convex optimization problem, this unconstrained bath fit can and often does get trapped in local minima, which strongly depend on the choice for the initial guess~\cite{mejuto-zaera_efficient_2020}. 
This calls for more efficient parameterizations of the bath that reduces the number of parameters to a smaller, motivated by symmetry or physics, number of parameters.

A key simplification, that we use in all the parameterizations discussed in this paper, is to {\it{partition the bath in a collection of smaller bath sets}}, where each set contains as many bath orbitals as there are impurity orbitals in the chosen implementation. In other words, the bath is formed by an integer number  $N_{\text{sets}}=N_{b}/N_{C}$ of copies of the cluster and the bath Hamiltonian is block diagonal in the space of such copies. Hence we introduce an internal label $\alpha $ for the fermions of the bath site $\alpha$ within the bath set $\lambda$, which are now destroyed by operators $a_{\alpha\lambda\sigma}$

\begin{eqnarray}
H_\text{imp}&=&H_C+\sum_{\lambda=1,\sigma}^{N_{\text{sets}}}\sum_{\alpha,\beta=1}^{N_C}h_{\alpha\beta}^{\lambda} a^\dagger_{\alpha\lambda\sigma} a^\dagga_{\beta\lambda\sigma} \nonumber\\&+& \sum_{\lambda=1,\sigma}^{N_{\text{sets}}}\sum_{\alpha,\beta=1}^{N_C}\left( V_{\alpha\beta}^{\lambda}a^\dagger_{\alpha\lambda\sigma} c^\dagga_{\beta\sigma} + H.c.\right)
\label{H_imp_sets}
\end{eqnarray}

and the hybridization function reads
\begin{equation}
 \Delta^{\text{Bath}} (i\omega)_{\alpha\beta} = \sum_\lambda^{N_{\text{sets}}}\sum_{\gamma,\gamma'=1}^{N_C}V_{\alpha\gamma}^{\lambda}({i\omega \mathbb{I}_{N_{C}}-\hat{h}_\lambda})^{-1}_{\gamma\gamma'}V_{\beta\gamma'}^{\lambda},
 \label{general_hyb_ansatz}
\end{equation}
where $\mathbb{I}_{N_{C}}$ is the identity matrix of dimension $N_C$.

We will now introduce and discuss in some details three ways to parameterize the bath which have been successful at improving the reliability of the fit ~\cite{koch_sum_2008,mejuto-zaera_efficient_2020,Capone2006-competition,Civelli2005}.
To illustrate this discussion, we also present a schematic comparison highlighting the differences between the three bath parameterization in the case of a 2-site cluster in Figure \ref{fig:bath_params_summ}. 

\paragraph{SDR:} In this case, we introduce a constraint in the bath parameterization that will allow us to perform the fitting step of the CDMFT procedure using semi-definite relaxation (SDR) optimization techniques. 
We begin by considering the case of a completely diagonal bath Hamiltonian  $h^\lambda_{\alpha,\beta}=\epsilon^\lambda_\alpha \delta_{\alpha\beta}$, where $\delta_{\alpha\beta}$ is the Kronecker delta. In this case, the expression for the hybridization function becomes:
\begin{equation}
\label{eq:Delta_eps_diag}
 \Delta^{\text{Bath}} (i\omega) = \sum_{\lambda}^{N_{\text{sets}}}\sum_\alpha^{N_C} \frac{\bm{V}_\alpha^\lambda\otimes {\bm{V}^\lambda_\alpha}^\dagger}{i\omega - \epsilon^\lambda_\alpha}
\end{equation}
where now, the $\bm{V}_\alpha^\lambda$ are vectors of dimension $N_C$, containing the couplings of bath site $\alpha$ in bath set $\lambda$ to the impurity degrees of freedom. 

Now, if we also add the constraint that all bath orbitals within each set have the same bath energy, i.e., 
$\epsilon^\lambda_\alpha = \epsilon_\lambda$, we can then relax the ansatz in Eq.~\eqref{general_hyb_ansatz} by replacing the sum of rank-1 matrices within a set 
$\sum_\alpha^{N_C}{\bm{V}_\alpha^\lambda\otimes {\bm{V}^\lambda_\alpha}}^\dagger$ by a positive semi-definite matrix $X_\lambda$ without rank restriction. 
So that, the hybridization function effectively becomes:
\begin{equation}
 \Delta^{\text{Bath}}_{\text{SDR}}(i\omega) = \sum_{\lambda}^{N_{\text{sets}}} \frac{X_\lambda}{i\omega - \epsilon_\lambda},
\end{equation}
allowing us to rewrite the optimization problem as finding the set of $\{\epsilon_\lambda\}_{\lambda=1}^{N_{\text{sets}}}$ that minimizes
\begin{equation}    \tilde{\mathcal{J}}\left(\{\epsilon_\lambda\}_{\lambda=1}^{N_{\text{sets}}}\right)=\min_{X_\lambda\succeq 0 }\mathcal{J}_{\text{SDR}}\left( \{ \epsilon_\lambda,X_\lambda\}_{\lambda=1}^{N_{\text{sets}}} \right).
\end{equation}
For a fixed set of bath energies, $\mathcal{J}_{\text{SDR}}$ is now a convex function which can be minimized under the constraint $X_\lambda\succeq0$ (i.e. each $X_\lambda$ matrix must be positive semi-definite) using conic solvers  ~\cite{andersen_implementing_2003,tutuncu_solving_2003}.  
Once an optimal set  $\{\epsilon_\lambda\}_{\lambda=1}^{N_{\text{sets}}}$ has been found, one may extract the hybridization vectors from the $X_\lambda$ matrices by solving an eigenvalue problem ~\cite{mejuto-zaera_efficient_2020}.

This technique allows us to keep the maximal amount of degrees of freedom in the couplings between the bath sites and the impurity orbitals during the fitting step of the CDMFT procedure. However, the imposed constraint on the bath energies may be too strong for calculation with a minimal number of bath sets. The number of parameters that need to be adjusted in the fit when the bath is parameterized in this way scales as $N_{\text{sets}}\left(1+N_{C}^2\right)$. As we will see, this parameterization is the one that allows for the biggest total number of free parameters in the fit. 

In Fig. \ref{fig:bath_param_SDR}, we show this bath parameterization in the $1\times2$ cluster for a single band model. As can be seen, all the degrees of freedom are kept in the couplings $V_{\lambda\alpha}$, while all the bath energies $\epsilon_\lambda$ are the same within the set $\lambda$.

\paragraph{Irreps:} In this implementation, we 
also take Eq.~\eqref{eq:Delta_eps_diag} as a starting point. However, we now exploit the symmetries of the cluster to reduce the number of bath degrees of freedom. In the absence of spontaneous symmetry breaking, the symmetries of the cluster must be reflected in the system’s observables, naturally constraining the form of the bath Hamiltonian.
In particular, both the Green's function and the hybridization function should be block-diagonal in a basis of impurity states that transform according to the irreducible representations (irreps) of the symmetry group.
For each irrep $I$, there exist $N_I$ linearly independent vectors in the Hilbert space of the impurity that transform as $I$, where $N_I$ corresponds to the dimension of the irrep. Enforcing block-diagonality in this symmetry-adapted basis significantly reduces the number of free parameters needed to fit the hybridization function~\cite{koch_sum_2008}.
We can thus write the hybridization function in this parameterization as:
\begin{equation}
 \Delta^{\text{Bath}}_{\text{Irrep}} (i\omega) = \sum_{\lambda}^{N_{\text{sets}}}\sum_I^{N_{\text{Irrep}}} \frac{\sum_\nu^{N_I} \lvert V_{I,\nu}^\lambda \rvert^2~ \bm{v}_{I,\nu} \otimes {\bm{v}_{I,\nu}^\dagger}}{i\omega - \epsilon^\lambda_I},
\end{equation}
where $\bm{v}_{I,\nu}$ are the $N_I$ linearly-independent constant vectors transforming according to the irrep $I$, and $N_{\text{Irrep}}$ is the total number of irreps.

Nevertheless, in the special case of an impurity model without interband hoppings, the couplings are expected to become diagonal in the band index. Under this condition, the bath can effectively be treated as if each bath set splits into $N_{\text{bands}}$ subsets. Each subset being coupled to only one of the bands, and containing only one bath orbital per each irrep of the point group of cluster.

Fig. \ref{fig:bath_param_Irrep} illustrates this in the case of the $1\times 2$ cluster for a single band model. In this simple case, the symmetry group is $C_2$, which only contains two irreducible representations: $A$ and $B$. We can thus build an orthogonal basis of states that transform as these irreps:

\begin{equation*}
\begin{split}
    \bm{v}_{A}=    &(+,+)\\
    \bm{v}_{B}=&(-,+)\\
\end{split}
\end{equation*}
where the cluster site labels are ordered from top to bottom.
Then, within each sets of baths, we assign to each bath orbital one of the irreducible representations of the cluster. In this case, we can then have two bath degrees of freedom per bath site, one for the bath on-site energy and one for its coupling to the impurities.

\paragraph{Symmetry-preserving Replicas:} Another way to reduce the number of parameters in the fit is to start from Eq.~\eqref{general_hyb_ansatz}, and 
build an internal bath Hamiltonian $\hat{h}_\lambda$ that contains only the same physical components as the impurity. Therefore, in this case, we will give each set of baths the same internal structure of the non-interacting part of $H_C$ (the impurity Hamiltonian) and add as many degrees of freedom as allowed by the symmetry of the cluster. Similarly, the bath couplings will only be allowed to appear in a way that respects the symmetries of the cluster impurity~\cite{Capone2004-1d,Civelli2005,Capone2006-competition,Civelli2008-TG,Kancharla2008}.

For example, in the case of the single-band $2\times 2$ cluster, the bath sites within each set $\lambda$ will have an on site energy $\epsilon_\lambda$, and nearest neighbor hopping $t_\lambda$, and a next-nearest neighbor hopping $t'_\lambda$. Analogously, the couplings will also be limited to local $V_{11}^\lambda$, nearest-neighbour $V_{12}^\lambda$ and  next-nearest-neighbour $V_{13}^\lambda$ kinds.

 In principle, this scheme can be generalized to allow for solution breaking the lattice symmetry. For instance, one could allow the hopping along the $x$ direction to be different from the $y$ direction, leading to bond-nematicity. In this work we limit however to symmetric solutions.

In this case, the number of degrees of freedom scales as $2N_{\text{sets}}\left(1+N_\text{hop}\right)$ where $N_\text{hop}$ is the number of independent hoppings included in the Replica description.
It turns out that by diagonalizing the resulting internal bath Hamiltonian of each set, we can show that the symmetry-preserving replica implementation can be described by an Irrep parameterization with some added constraints (see Appendix \ref{Replica_in_Irrep_basis}).

See Fig. \ref{fig:bath_param_Replica}, for an illustration of this bath parameterization in the case of the $1\times2$ cluster. In this minimal case, the internal structure of the Replica is the same as the one in the impurity, and the couplings are therefore also limited to only two different kinds. 

Each parameterization offers distinct advantages: the replica method is the easiest to implement in a generic situation, allowing for the addition of intuitive, physically motivated constraints and yielding results that are simple to interpret. On the other hand, the irrep decomposition provides the most systematic framework to preserve all the relevant symmetries. Finally, the SDR approach provides the numerically most robust exploration of the parameter landscape of the fit,  mitigating issues related to local minima and fitting stability.

\begin{figure*}
    \centering
    \includegraphics[width=\linewidth]{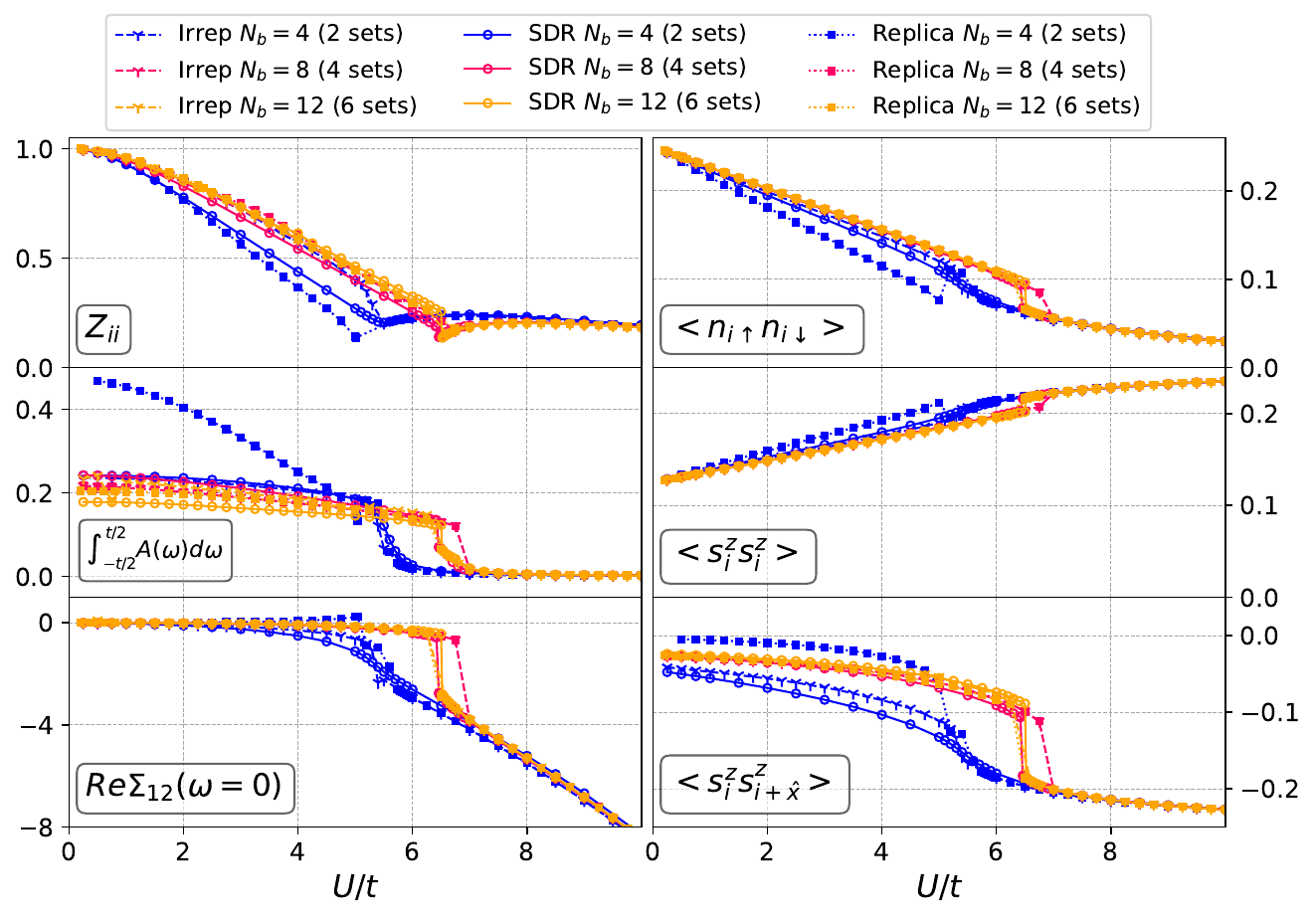}
    \caption{Cluster quantities of the $1\times 2$ CDMFT solutions of the half-filled single band Hubbard model as a function of the interaction strength. \textbf{Left}: Quasiparticle residue of the impurity, spectral weight at low frequency in the local spectral function, and real part of the off-diagonal self energy at zero frequency. \textbf{Right}: Double occupancy, local and nearest neighbor spin-spin correlations.}
    \label{fig:1x2x1 cluster observables}
\end{figure*}
\section{Results}
\subsection{SINGLE BAND \texorpdfstring{$1\times 2$}{1x2} CLUSTER}

\begin{figure*}
    \centering
    \includegraphics[width=1.00\linewidth]{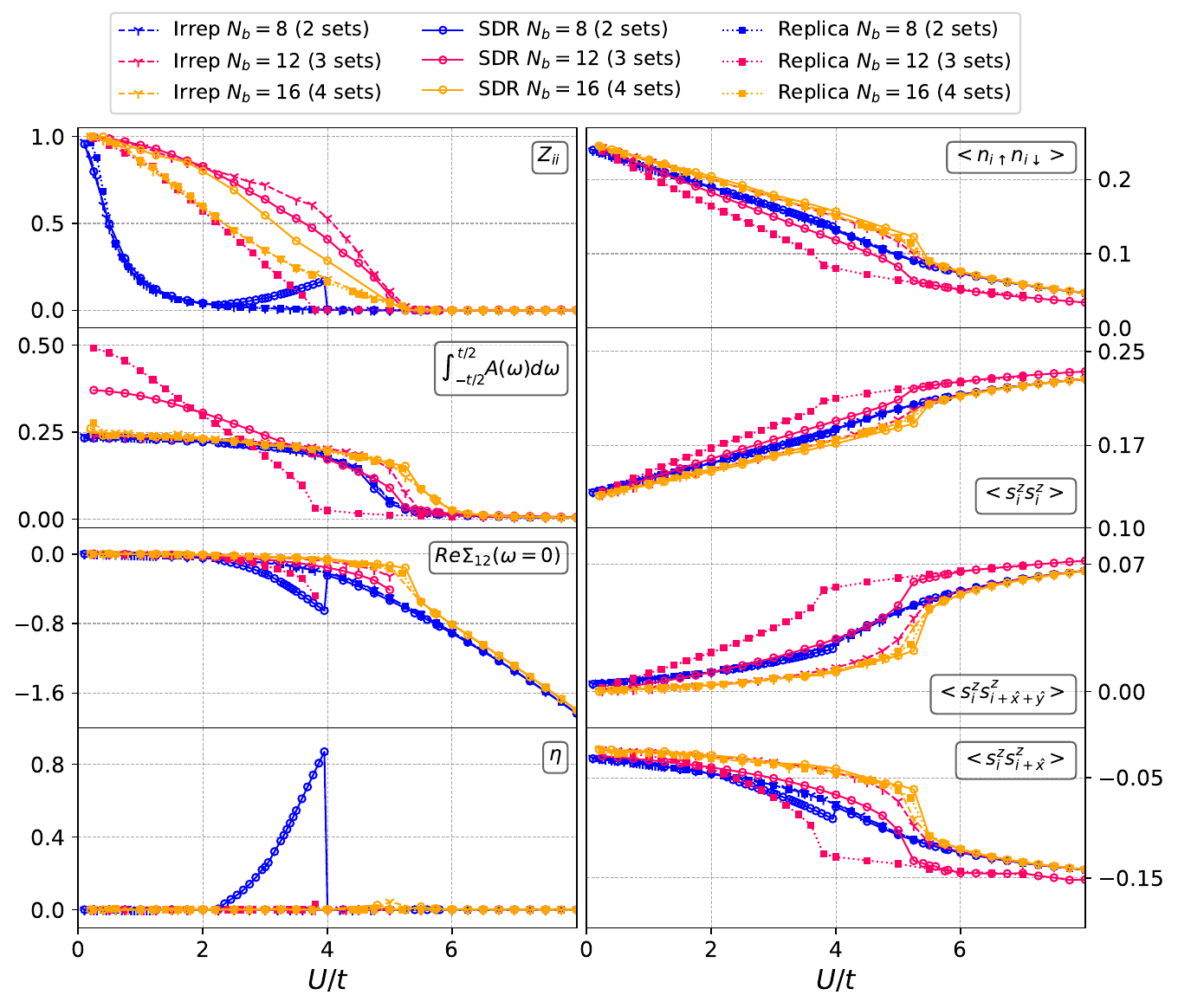}
    \caption{Cluster quantities of the $2\times 2$ CDMFT solutions of the half-filled single band Hubbard model as a function of the interaction strength. \textbf{Left}: Quasiparticle residue of the impurity, spectral weight at low frequency in the local spectral function, real part of the off-diagonal self energy at zero frequency and nematic order parameter. \textbf{Right}: double occupancy, local, next-nearest neighbor and nearest neighbor spin-spin correlations.}
    \label{fig:2x2x1 cluster observables}
\end{figure*}
\begin{figure}
    \centering
    \includegraphics[width=\linewidth]{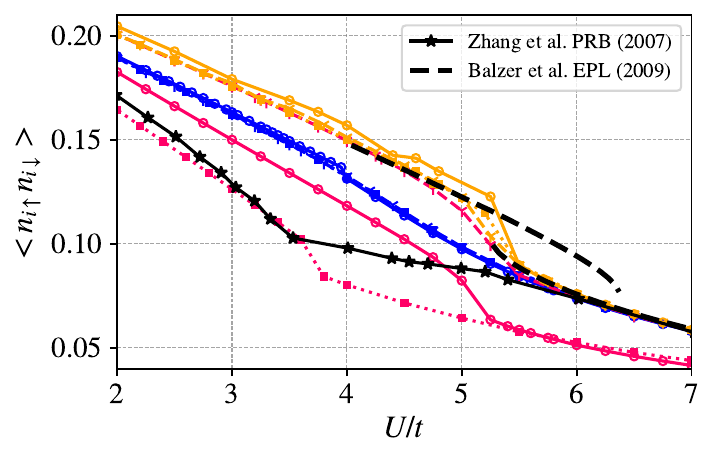}
    \caption{Comparison of our results for the double occupancies in the cluster sites with references ~\cite{zhang_pseudogap_2007} and ~\cite{balzer_first-order_2009} for the half-filled single band Hubbard model solved with a $2\times2$ impurity cluster.}
    \label{fig:docc_comp}
\end{figure}
A two-site cluster, or dimer, is the first and computationally cheapest implementation of cDMFT. In this system, we can increase the number of sets (copies) at feasible computational cost because the total number of orbitals grows slowly with the number of bath sets. This makes it an ideal system for studying the convergence of results across the different bath parameterizations as the number of bath orbitals is increased. Notice, however, that the dimer clearly has a lower symmetry than the square lattice. In this work we use the most natural tiling, in which the two-dimensional lattice is reconstructed translating the cluster of multiples of the vectors $(2,0)$ and $(0,1)$, assuming that the cluster is aligned along the $x$ axis. This choice clearly favors one-dimensional correlations along the specific direction defined by the dimer, which can introduce biases in the computed quantities of the system.

In Figure \ref{fig:1x2x1 cluster observables}, we present various cluster quantities computed as a function of the interaction strength, employing different bath parameterizations and sizes. 
We point out that at half-filling the parity of $N_{\text{sets}}$ plays an important role, similarly to single-site DMFT.
To maintain a particle-hole symmetric Hamiltonian, which is a necessary and sufficient condition for half-filling for our model, pairs of sets with opposite energies must be present. If the number of sets is odd, this implies that one level with zero energy must be included.

This constraint inherently favors metallic solutions which are defined by a finite weight at zero energy. Moreover, insulating solutions require a vanishing hybridization for the zero-energy baths, which implies that they are disconnected from the impurity. This can be hard to obtain numerically, but it would also imply that the results will coincide with those with $N_{\text{sets}} -1$.
Therefore, for this cluster geometry where we can easily afford it, we have chosen to focus exclusively on an even number of bath sets.

In particular, we used ED for all the calculations performed using $N_b=4$ and $N_b=8$, while, for $N_b=12$, we used ASCI with $\sim 2.5\times 10^4$ as the typical number of determinants.
At the smallest, nontrivial bath size possible $N_{\text{sets}}=2$, we observe differences between the three parameterizations for all the computed cluster quantities, but as we increase the number of sets, these differences become noticeable only around the critical $U$ for $N_{\text{sets}}=4$, and virtually disappear for $N_{\text{sets}}=6$. Thus, we can say that the different bath parameterizations converge to the same solution once the bath becomes big enough. 

This value of $N_{\text{sets}}$ is however challenging to achieve in CDMFT calculations with larger clusters and/or more than one orbital. For instance, even for a $2\times 2$ plaquette with a single band, $N_{\text{sets}}=6$ corresponds to a bath of 24 orbitals, resulting in a 
system size that is computationally  expensive even for the ASCI solver.

As seen from the sharp drop in spectral weight at low frequency (middle left panel in Figure \ref{fig:1x2x1 cluster observables}), the solution found with $N_{\text{sets}}=4$ and $N_{\text{sets}}=6$ shows a clear opening of a gap at $U_c\approx6.5t$. This critical value for the interaction strength is in good quantitative agreement with previous results of a dimer cluster DMFT geometries both at zero temperature and at finite temperature ~\cite{zhang_pseudogap_2007,meixner_mott_2024}.
At $U > U_c$, we observe no divergence of the imaginary part of the self-energy in the imaginary frequency axis, leading to a finite quasiparticle weight even after the transition (top left panel in Figure \ref{fig:1x2x1 cluster observables}). The gap opening coincides with a discontinuity, followed by a rapid increase, in the absolute value of the real part of the off-diagonal self energy (bottom left panel in Figure \ref{fig:1x2x1 cluster observables}). It is therefore the development of non-local correlation that drives the system insulating, in contrast to single-site DMFT where there is a divergence of the local self-energy in the Mott phase. This is in agreement with analogous calculations carried out at low but finite temperature in the $1\times 2$ cluster ~\cite{meixner_mott_2024}.
The discontinuity in the transition can also be seen from the double occupancy and the spin-spin correlations (right panels in Fig. \ref{fig:1x2x1 cluster observables}). All these quantities  asymptotically approach the values we would expect for an isolated singlet state at high values of the interaction. In particular, the value of the nearest-neighbor spin correlation is always negative, pointing to the presence of antiferromagnetic tendencies for all values of the interaction, and presents the biggest jump among the computed observables at the MIT. 

Notice that the solution with the smallest number of baths shows in general a good qualitative agreement at small and large values of the interaction. The biggest difference between this solution and the large bath solutions appears in the description of the MIT transition, at intermediate values of the interaction. Mainly, the transition is closer to a smooth crossover between the two phases, specially for the solutions found using the SDR and Irrep bath parameterization, with the gap opening at smaller $U_c$.

\begin{figure*}
    \centering
    \includegraphics[width=0.9\textwidth]{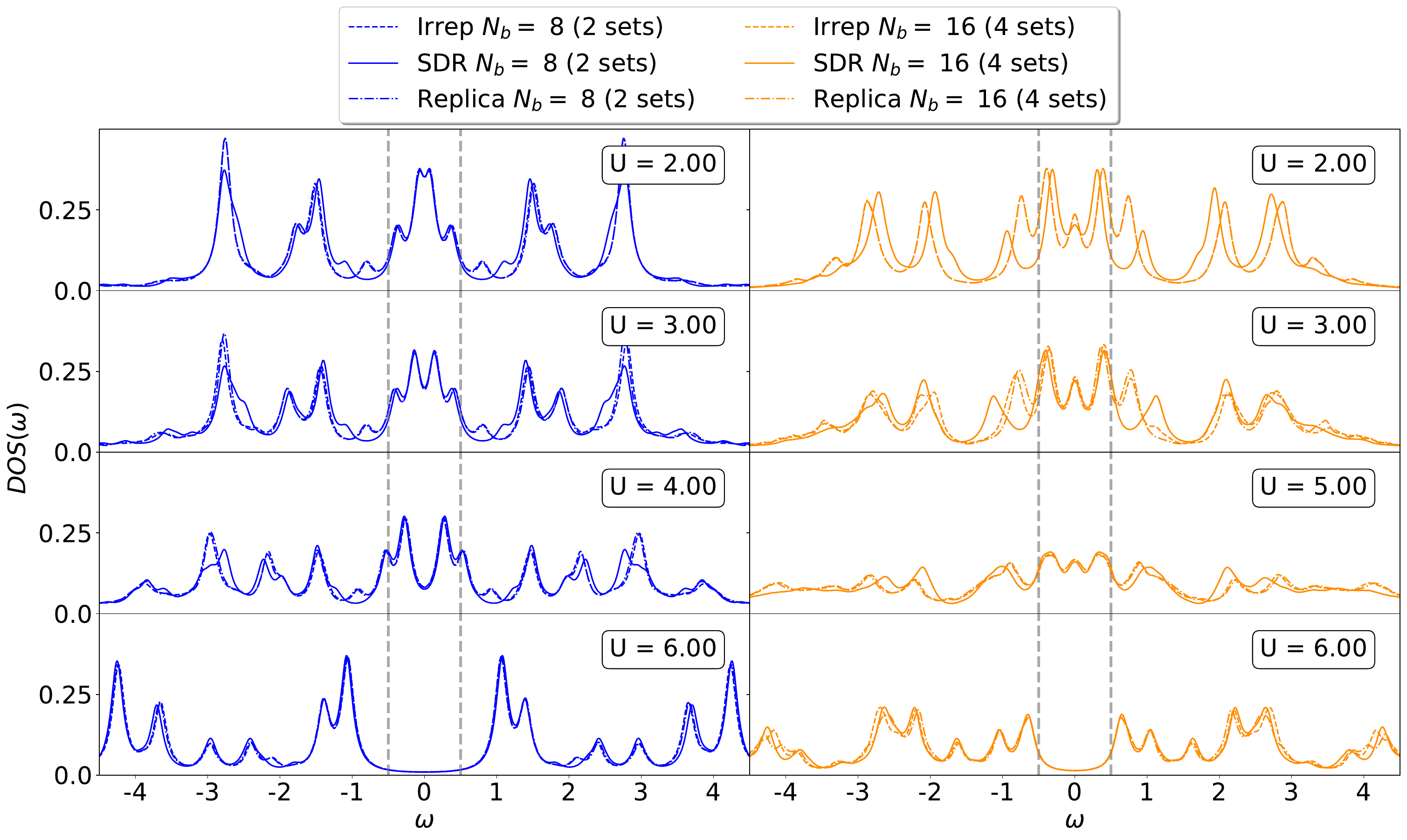}
    \caption{Evolution of the density of states with increasing interaction strength in \textbf{(a)} the $N_{b}=8$ solution and \textbf{(b)} the $N_{b}=16$ solution. The dotted lines show the integration limits $\left[-\frac{t}{2},\frac{t}{2}\right]$ for the estimation of spectral weight at low frequency.}
    \label{fig:2x2x1_spectral}
\end{figure*}

\subsection{SINGLE BAND \texorpdfstring{$2\times 2$}{2x2} CLUSTER}
\label{sec:1b2x2}

We now turn to the results obtained using a $2\times 2$ plaquette, which respects all point-group symmetries of the original model under investigation.
In what follows, we label the cluster sites from left to right and from top to bottom, assigning the numbers 1 through 4, as illustrated in Fig. \ref{fig:2x2_convention}.
In this section, we discuss the evolution of both cluster and lattice observables as a function of interaction strength, computed using various bath parameterizations and sizes.

For these calculations, we used ED in the case of $N_b=8$, and ASCI for $N_b=12$ and $N_b=16$. The typical number of determinants used for the $N_b=12$ computations was $\sim 2\times10^5$ determinants, while, for $N_b=16$, we kept $\sim 5\times 10^5$ determinants.

\subsubsection{Cluster Quantities}

Also for the $2\times2$ plaquette, we find noticeable differences in the results obtained for different bath sizes in the intermediate interaction regime. The solutions with the smallest nontrivial number of baths, $N_{b}=8$ with the replica and irrep bath parameterizations, show a smooth crossover between a metallic and an insulating solution with a rapidly decaying quasiparticle weight at weak values of the interaction strength (blue curves in the top left panel in Figure \ref{fig:2x2x1 cluster observables}).

Instead, for the same $N_{b}=8$, the SDR, which is the most constrained bath parameterization for small $N_{\text{sets}}$, converges to a different solution before the opening of the gap. This region is characterized by an increase of the quasiparticle weight and a breaking of the rotational symmetry between the $x$- and $y$-direction, leading to a nematic metallic solution.  
Measuring the degree of nematicity as:
\begin{equation}
    \eta = |\text{Re} \Sigma_{12}(\omega=0)-\text{Re}\Sigma_{13}(\omega=0)|,
\end{equation}
where $\Sigma_{12}$ and $\Sigma_{13}$ refer to the $x$- and $y$- direction of our two-dimensional lattice, we find an increasing trend as a function of $U$ that follows the growth of $Z$.
Interestingly, the enhancement of the off-diagonal part of the self-energy associated with the development of nematicity is not accompanied by the development of a gap because of the direction dependence of the self-energies.
We also notice that solutions of this kind, where there is a differentiation along the $x$ and $y$ directions, would not be allowed by construction on the symmetry preserving replica structure for the baths (see Appendix \ref{Replica constraint in irrep basis}), while the irrep bath structure, which could potentially allow for this sort of solution in this system, converged to non-nematic solutions even when imposing a nematic starting point for the bath.
This suggests that the nematic solution is only stablized in SDR because this scheme frustrates the standard metallic solution, which remains the most stable solution.
Yet, we can rationalize the appearance of the nematic region as a possible sign of the proximity of the ground state of the half-filled 2D Hubbard model to nematic excited states, which may define the new ground state by slight changes in the parameters of the model ~\cite{okamoto_dynamical_2010,su_coexistence_2011,fang_local_2013,kaczmarczyk_coexistence_2016}.

With $N_{b}=16$, we find again a very good agreement between the different bath parameterizations for all observables but the quasiparticle weights. The picture that we recover from these solutions is qualitatively similar to the one we got from the calculation with the dimer impurity. The spectral weight at low frequency decreases very slowly for intermediate values of the interaction, until a sharp drop is observed at a critical interaction strength. At this point, discontinuous jumps are seen in the spin correlation functions and the double occupancies, along with an important enhancement of the real part of the nearest neighbor component of the cluster self-energy. 
This corresponds to a first order phase transition at $T=0$, at variance with the previous CDMFT on the single band 2D Hubbard model in Ref.~\cite{zhang_pseudogap_2007}, while variational cluster approximation (VCA) method~\cite{balzer_first-order_2009} finds a discontinuous transition. Indeed, in Fig. \ref{fig:docc_comp} we can see that the double occupancies from the VCA study match the ones obtained in our $N_{b}=16$ solutions. On the other hand, the double occupancies reported in the previous CDMFT study~\cite{zhang_pseudogap_2007} show a similar behavior to our solutions with $N_{b} > 8$ for low values of the interaction and very good agreement with our even $N_{\text{sets}}$ solutions at strong interaction, but have a very different behavior around the phase transition.

The solutions found with $N_{b}=12$ (which corresponds to an odd number of sets) also display a discontinuous transition,  but the considerable quantitative differences between the solutions found with different bath parameterization suggest that here finite-size effects are stronger, making the outcome less trustworthy, as expected for an odd number of sets.

Notice that, in the $N_b=16$ case, we find  a smaller value for the critical interaction strength $U_c\approx5.5t$ compared to the one obtained in the calculation with the dimer impurity. This reduction is in agreement with the general trend that in larger impurity clusters, which allow for longer range AFM correlations, the critical value for interaction is reduced ~\cite{meixner_mott_2024}. 
This value for the critical interaction strength is also in great quantitative agreement with the values reported in the VCA study at $T=0$ and at finite temperature in other CDMFT calculations ~\cite{park_cluster_2008,liebsch_multisite_2008,meixner_mott_2024}.

Additionally, as for the dimer simulations, we find that the results for double occupancies and spin-spin correlation functions  are much less dependent on bath size and parameterizations than quasiparticle properties and self-energies, at least for small values of the interaction and in the insulating state. 
However, for the plaquette, the nearest-neighbor and next-nearest-neighbor spin correlations do not converge to those of a bond singlet state even at the highest values of the interaction. This reflects the fact that the AFM correlations in the plaquette become different from those of a single site, in contrast with the dimer.

In order to complete the picture that we get from the cluster observables, we finally turn to the local spectral functions shown in Fig. \ref{fig:2x2x1_spectral}. 
We have chosen to plot the spectral functions at values of the interaction where the quasiparticles of the system are still fairly coherent  ($U=2t$ and $U=3t$ for both solutions), just before the metal-insulator transition ($U=4t$ for $N_b=8$ and $U=5t$ for $N_b=16$), and where the system is clearly insulating ($U=6t$ for both solutions).
In the $N_{b}=8$ solution, the appearance of a twin peak structure is observed at $U\approx1.5t$. These twin peaks are continuously separated as $U$ increases, until the solution becomes gaped at $U\approx5t$. This behavior is very similar to what was reported in previous exact diagonalization CDMFT studies with $N_{b}=8$~\cite{kyung_pseudogap_2006,zhang_pseudogap_2007} for intermediate $U$, while a single quasiparticle peak was reported to persist up to $U=4t$.
Good quantitative agreement was found instead with the reported cluster self-energies at $U\geq4t$ from another ED-CDMFT study using $N_{b}=8$ ~\cite{zgid_truncated_2012}.

On the other hand,the $N_{b}=16$ solution shows a three-peak structure, with a central peak always at zero, and side peaks which now remain at fixed energy despite the increasing interaction strength. Instead, the whole central structure transfer spectral weight to higher frequencies as the the system becomes more strongly correlated, until a gap opens abruptly at $Uc$.
Remarkably, a very similar triple peak structure was reported in another VCA calculation at half-filling ~\cite{schafer_fate_2015}, and in a previous CDMFT using finite-temperature exact diagonalization with a $2\times2$ impurity and only 8 bath sites, but where the DMFT self-consistent procedure  was entirely performed in the cluster momentum basis, where the self-energy is diagonal ~\cite{liebsch_multisite_2008}.

Notice that for both sizes of the bath, we find a good convergence with respect to the different bath parameterizations in this case. In the $N_{b}=8$ case the local spectral function are remarkably similar even at $U=3t$, where the SDR solution is nematic. However, as we will discuss later, this will no longer hold for the $k$-resolved spectral functions.
\begin{figure*}
\centering
\includegraphics[width=0.75\textwidth]{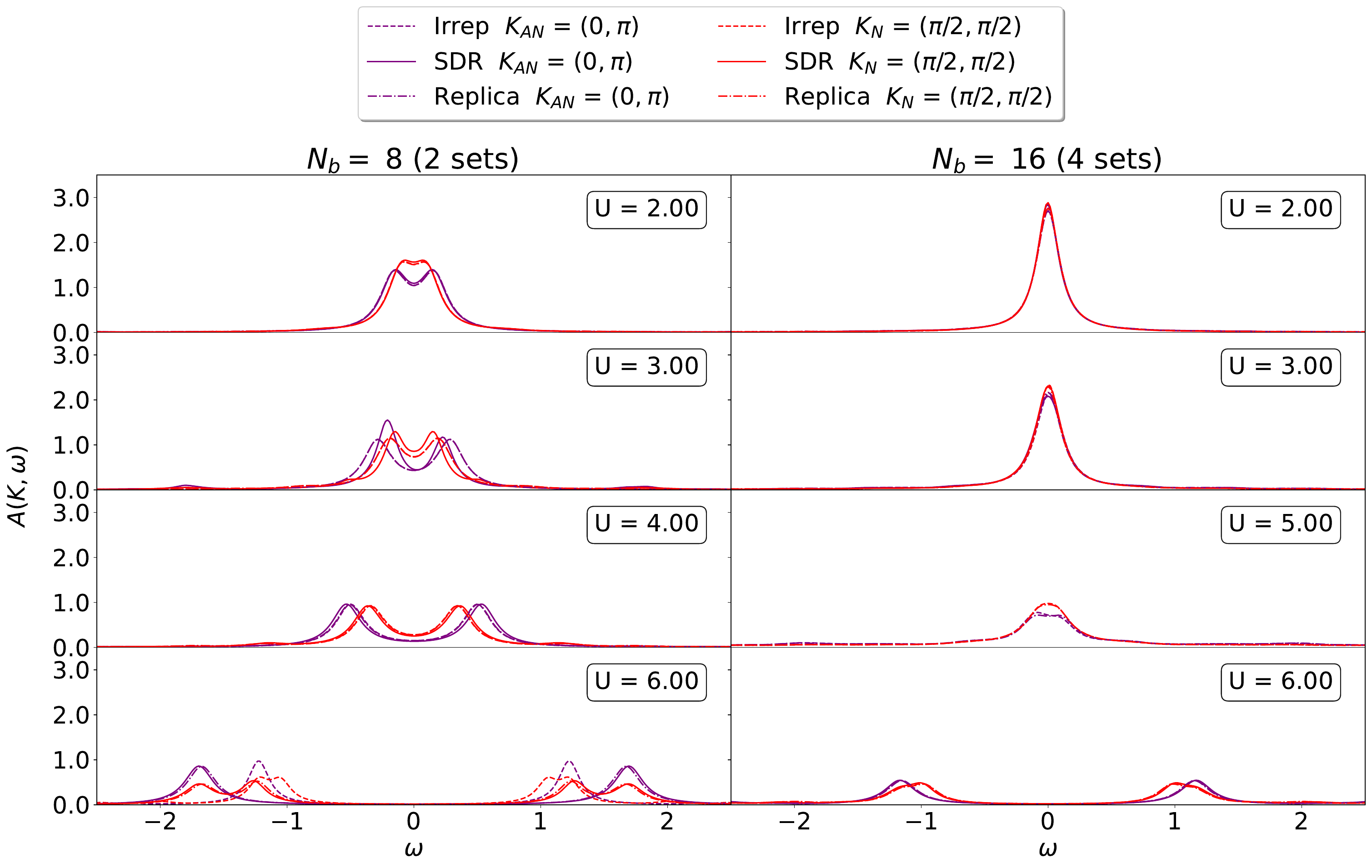}
    \caption{Change in the spectral function at the node ($K_N=X=(0,\pi)$)and the antinode ($K_{AN}=M/2=(\pi/2,\pi/2)$ ) points in the Brillouin Zone with increase interaction strength in the half-filled single band Hubbard model solved with $2\times2$ impurity cluster. \textbf{(a)} Solution with $N_{b}=8$  \textbf{(b)} Solution with $N_{b}=16$. }
    \label{fig:2x2x1_nb8_AN_Nspectral}
\end{figure*}
\begin{figure*}
    \centering
    
    \raisebox{0.08\textwidth}{\small {$\bm{N_{b}=8}$}} 
    \hfill
    \subfloat{\includegraphics[width=0.29504\textwidth]{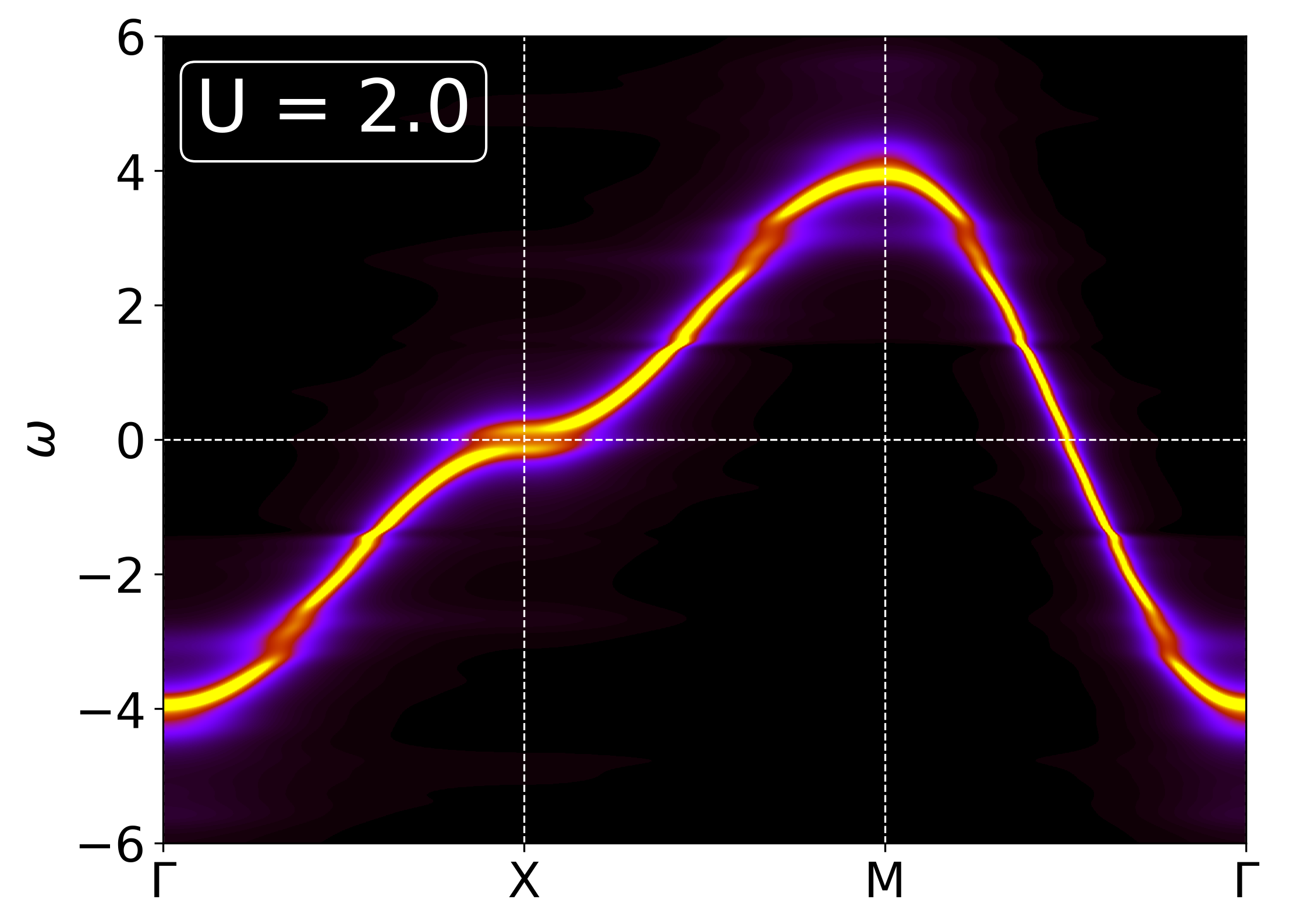}} 
    \subfloat{\includegraphics[width=0.29504\textwidth]{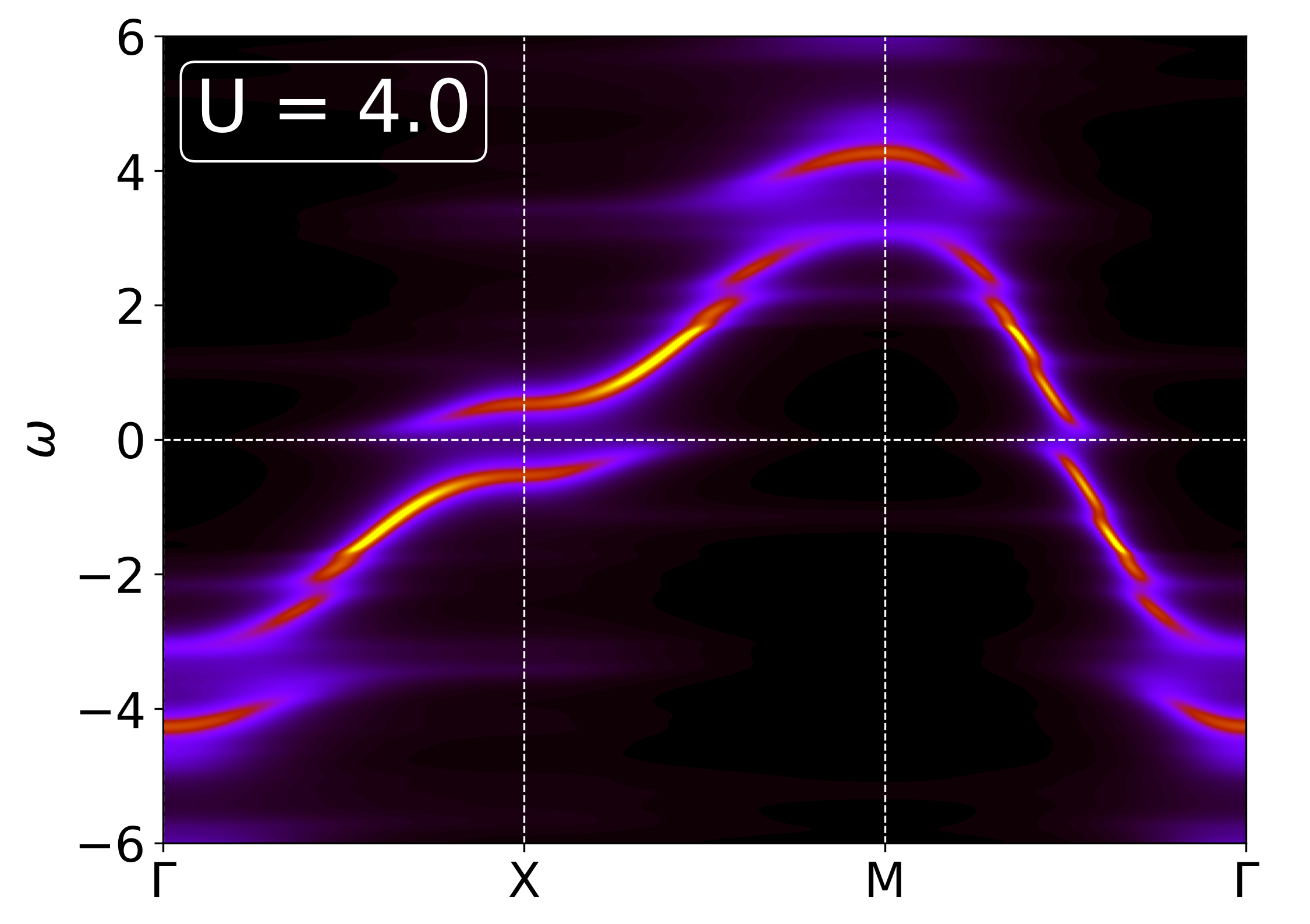}} 
    \subfloat{\includegraphics[width=0.32992\textwidth]{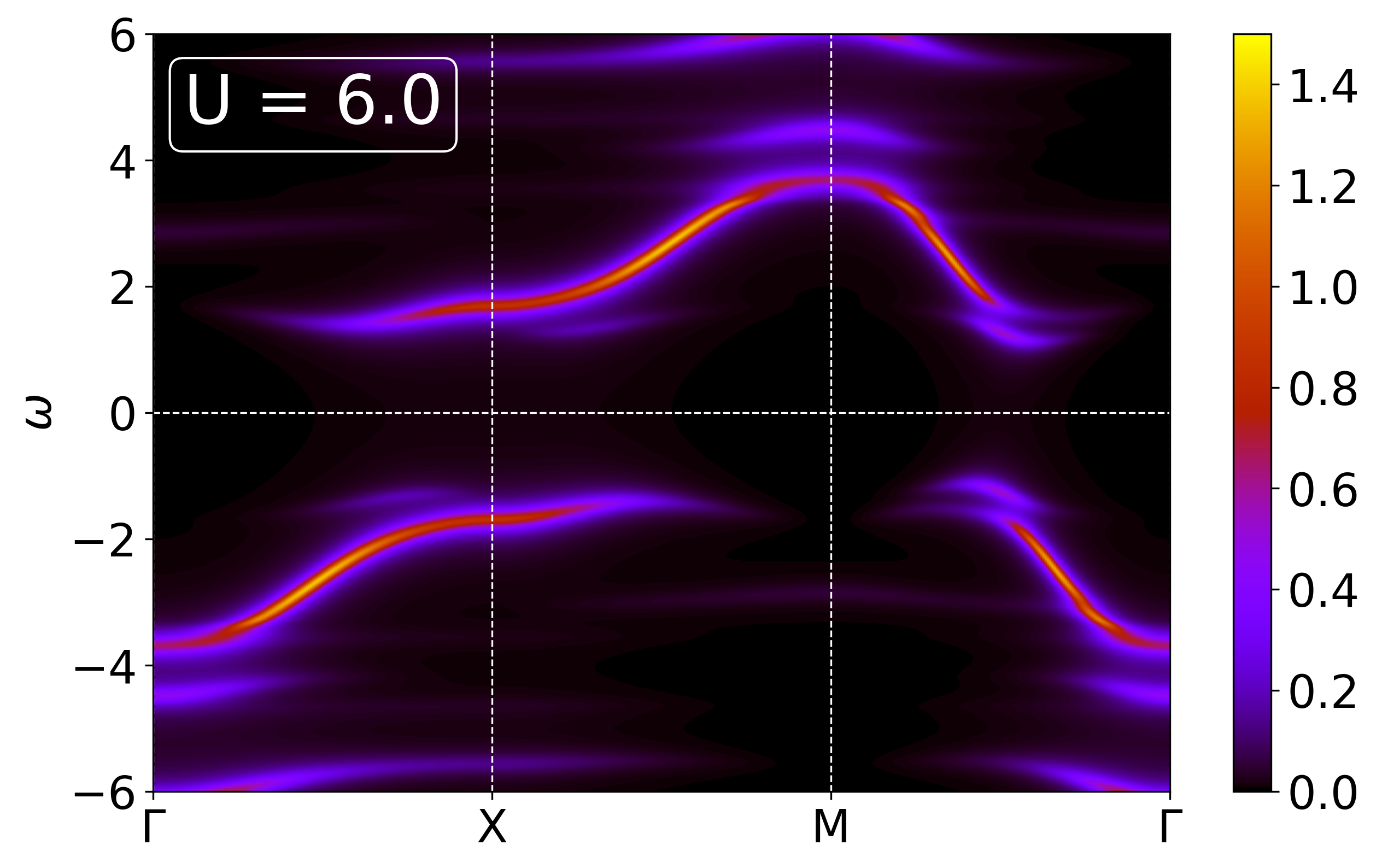}} \\
    \raisebox{0.08\textwidth}{\small $\bm{N_{b}=16}$} 
    \hfill
    \subfloat{\includegraphics[width=0.29504\textwidth]{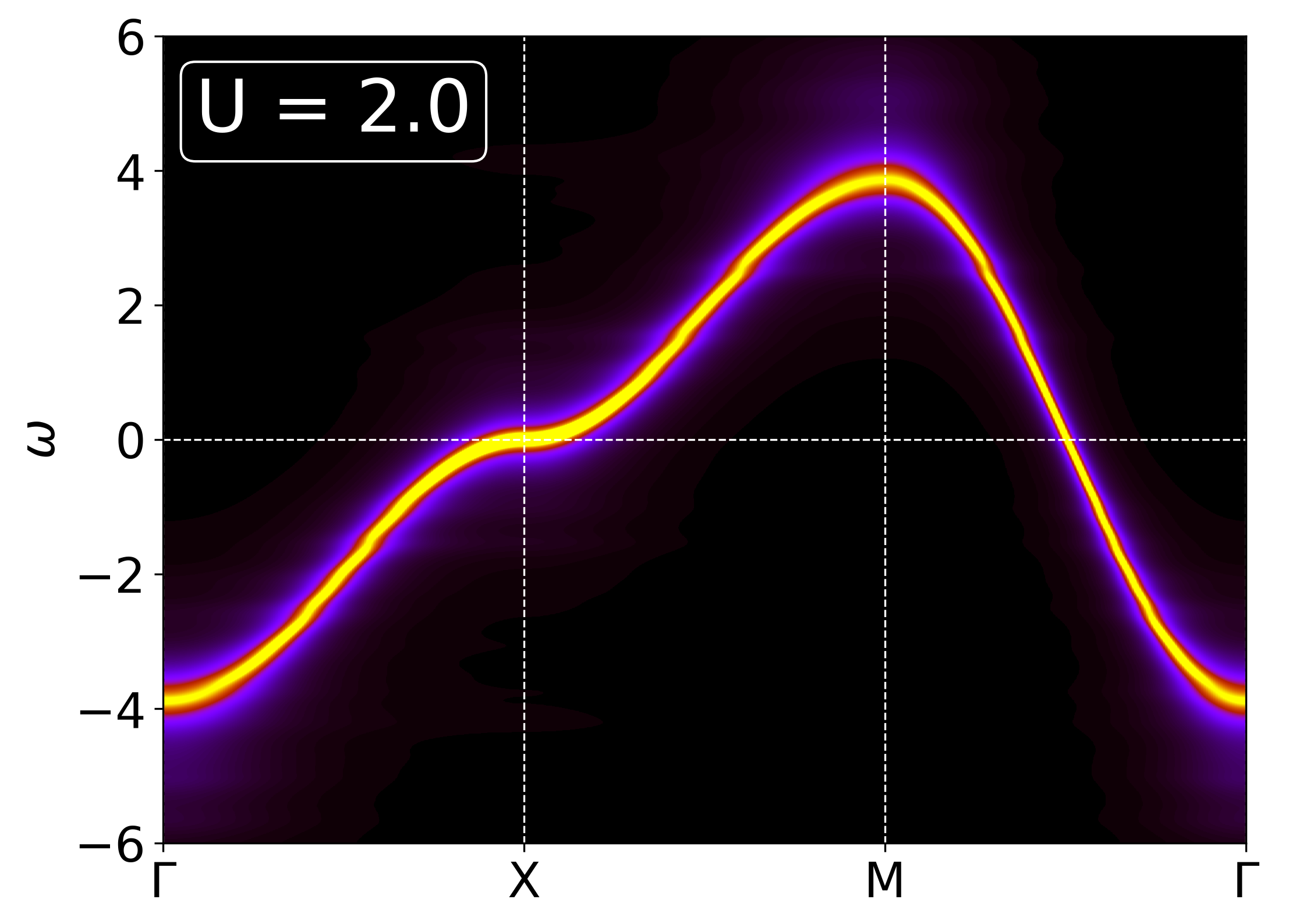}} 
    \subfloat{\includegraphics[width=0.29504\textwidth]{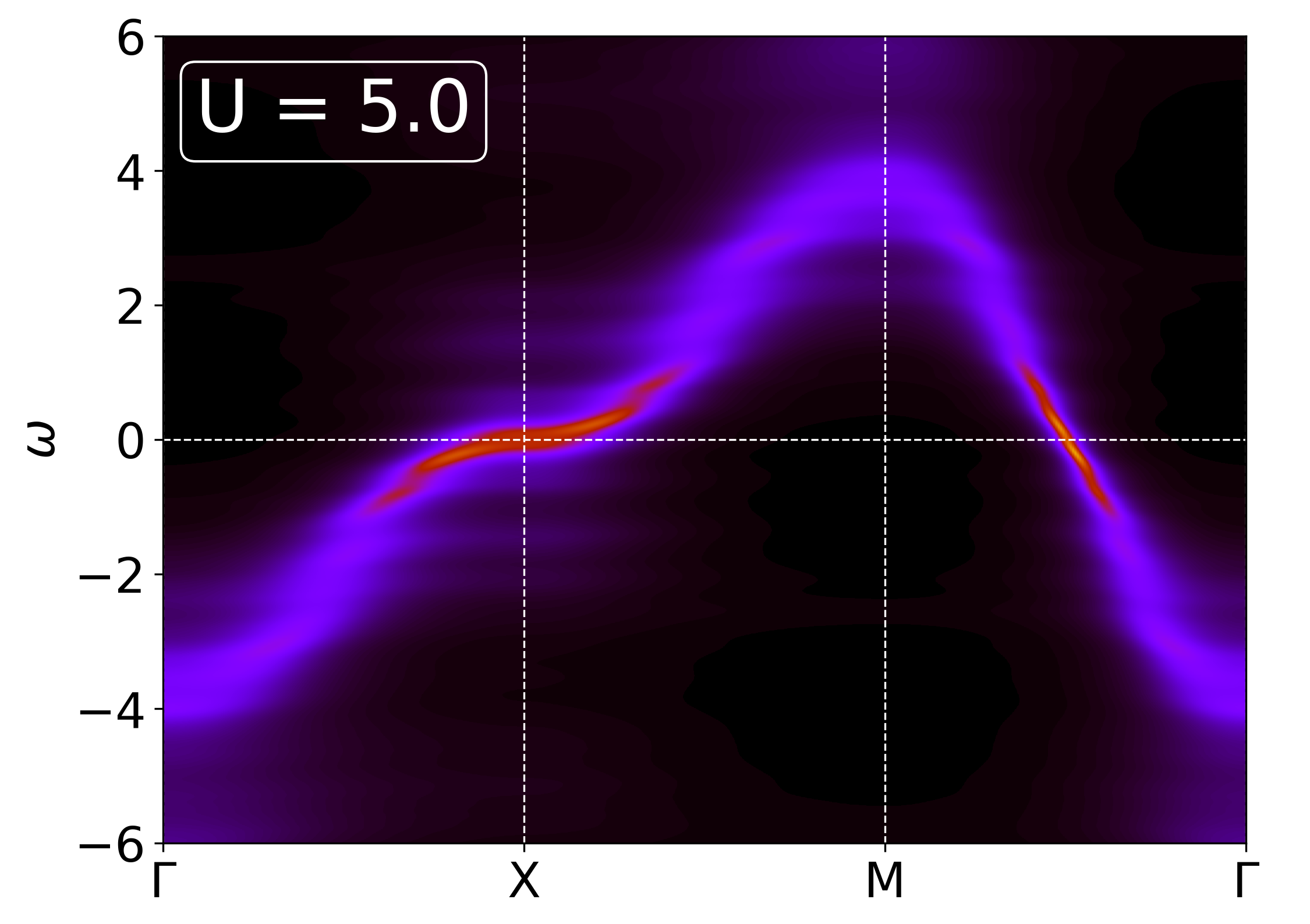}} 
    \subfloat{\includegraphics[width=0.32992\textwidth]{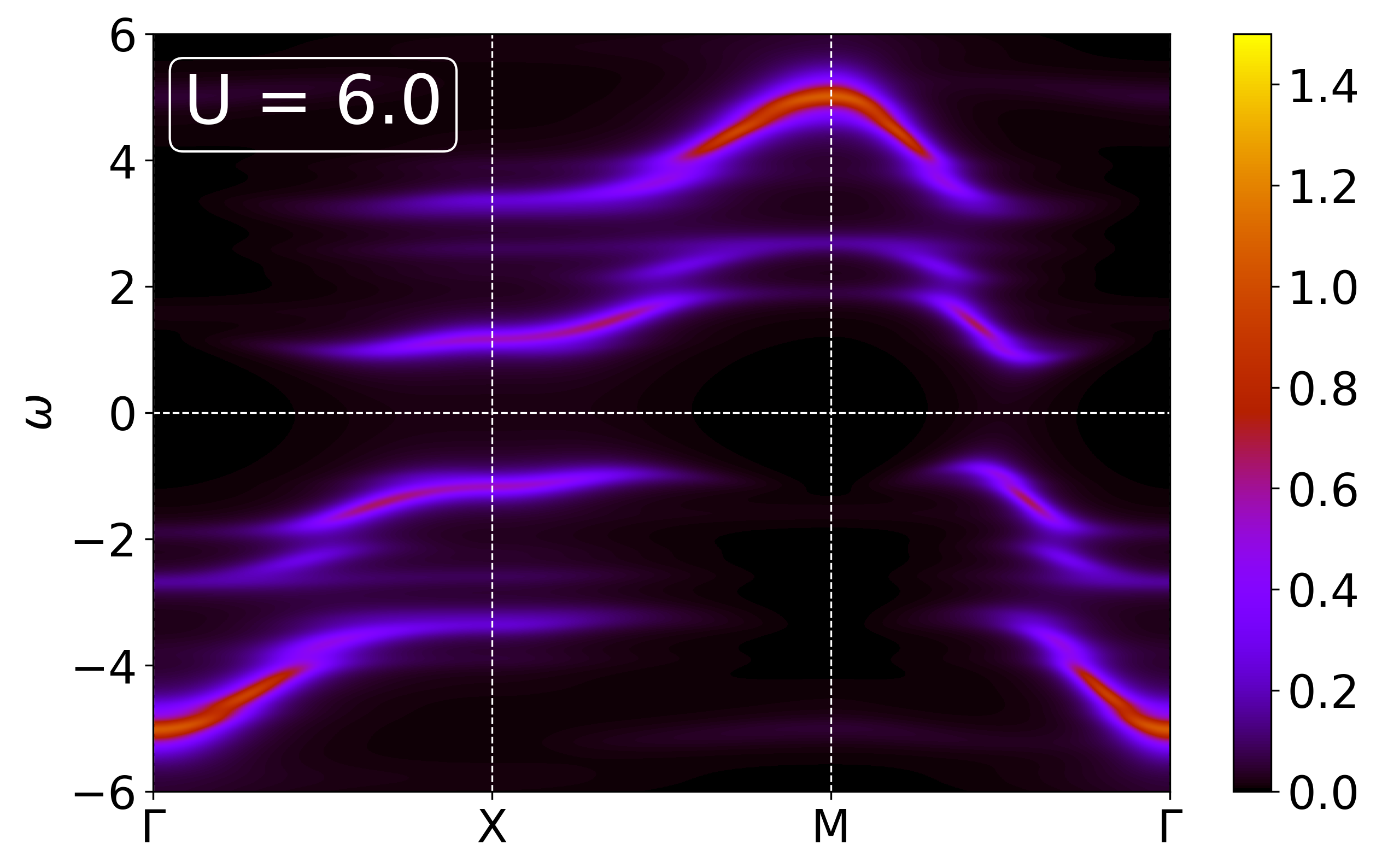}} \\
    
    \caption{Spectral functions for the half-filled single Hubbard model solved using CDMFT with a $2\times2$ impurity cluster computed and the Replica bath parameterization.}
    \label{fig:2x2x1_big_spectral_fig}
\end{figure*}
\subsubsection{Lattice quantities}
\label{Sec:lattice}
To complete our characterization of the plaquette solutions, we present some results for the momentum-resolved spectral properties. As it has been widely discussed~\cite{Civelli2005,Civelli2008-TG,stanescu_fermi_2006,Capone2006-competition,sakai_cluster-size_2012}, different periodization schemes can be used to extract this information. The choice essentially boils down to selecting a dynamical quantity $Q$ (Self-energy, Green's function, cumulant, \ldots) to be periodized according to
\begin{equation}
    Q(\bm{k},i\omega_n)=\frac{1}{N_\text{sites}}\sum_{ij}^{N_\text{sites}} Q_{ij}(i\omega_n)e^{i\bm{k}\cdot\left(\bm{r}_i-\bm{r}_j\right)},
\end{equation}
and then we compute the other quantities from $Q(\bm{k},i\omega_n)$. The choice is inspired by the nature of the functions. In particular, according to the parameters range, it is convenient to periodize the least singular quantity, i.e., the one that does not have poles. 
Therefore, at small to intermediate interactions $U < 6t$, we periodize the self-energy, while for $U\geq6t$, where the self-energy can become singular, we periodize the cumulant 
$M=\left[ i\omega_n+\mu-\Sigma\right]^{-1}$ which is expected to be a smooth function.
Moreover, the M-periodization scheme has been shown to perform better than periodization of the self-energy or the Green's function for insulating solutions or strongly correlated metallic solutions away from half-filling ~\cite{stanescu_cellular_2006,sakai_cluster-size_2012}.

Fig. \ref{fig:2x2x1_nb8_AN_Nspectral} displays the spectral function $A(k,\omega)$ computed at the node $K_N=X=(0,\pi)$ and the antinode $K_{AN}=M/2=(\pi/2,\pi/2)$ .
For $N_{b}=8$ the momentum-resolved spectral functions exhibit similar behavior to the local one, with twin peaks forming and continuously evolving into a gap as the peak separation increases with the interaction strength. In contrast, the $N_{b}=16$ solution shows a well defined single peak that slowly loses spectral weight at low frequency.

In both cases, we observe a small but noticeable momentum differentiation at intermediate values of the interaction, with the spectral weight at the antinode depleting faster than at the node.
This effect is consistent with the expected influence of antiferromagnetic (AFM) fluctuations within the impurity cluster. 

However, in the $N_b=16$ calculation, the abrupt opening of the gap happens at the same interaction strength for both points in the Brillouin zone. This reduction of the node-antinode dichotomy at high values of the interaction, right before the Mott transition, may indicate that local electronic correlations prevail over spatial fluctuations in this regime. This result aligns with diagrammatic non-local extensions of DMFT ~\cite{chatzieleftheriou_local_2024}.
In addition, we observe a breaking of particle-hole symmetry in the nematic regime when using the SDR bath parameterization with $N_b=8$. Specifically, at $U=3t$, the spectral function at $K=(0,\pi)$ exhibits a higher spectral weight at negative frequencies than at positive ones, while the opposite behavior is seen at $K=(\pi,0)$. Importantly, this symmetry breaking occurs in such a way that the local spectral function—obtained by summing the momentum-resolved spectral functions over $k$—remains particle-hole symmetric, as shown in Fig.~\ref{fig:2x2x1_spectral}.

Finally, Figure \ref{fig:2x2x1_big_spectral_fig} presents the full $k$-dependent spectral function along a closed path connecting high-symmetry points of the Brillouin zone of the square lattice. Here we focus on the solutions obtained using the replica bath parameterization. 
For $N_{b}=8$ at $U=2t$, the pseudogap structure is evident, with the twin-peak feature clearly visible at X while the quasiparticle peak remains coherent at M/2. At $U=4t$, a clear gap forms, accompanied by a broad structure at $\Gamma$ and M, which further evolves into two distinct peaks at $U=6t$.
In contrast, the $N_{b}=16$ solution exhibits a coherent spectral weight along the entire non-interacting dispersion at $U=2t$. At $U=5t$, a slight differentiation emerges between X and M/2, with these points retaining most of the spectral weight. Then, for $U=6t$, after the gap has opened, two additional bands form at X and M/2, while the majority of the spectral weight shifts to $\Gamma$ and M.

\subsection{TWO-BAND \texorpdfstring{$1\times 2$}{1x2} CLUSTER}
\begin{figure*}
    \centering
    \includegraphics[width=0.25\linewidth]{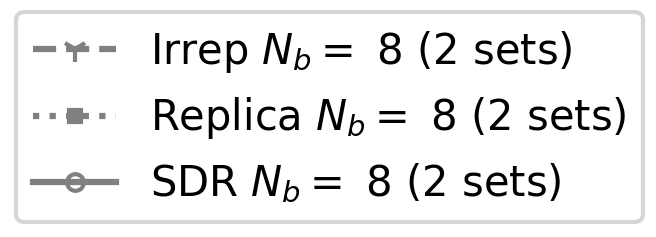}
    \includegraphics[width=0.12\linewidth]{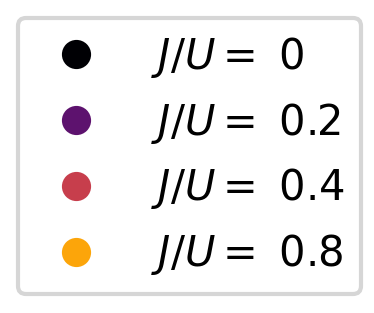}
    \begin{subfigure}{\textwidth}
    \subfloat{\includegraphics[width=0.9\linewidth]{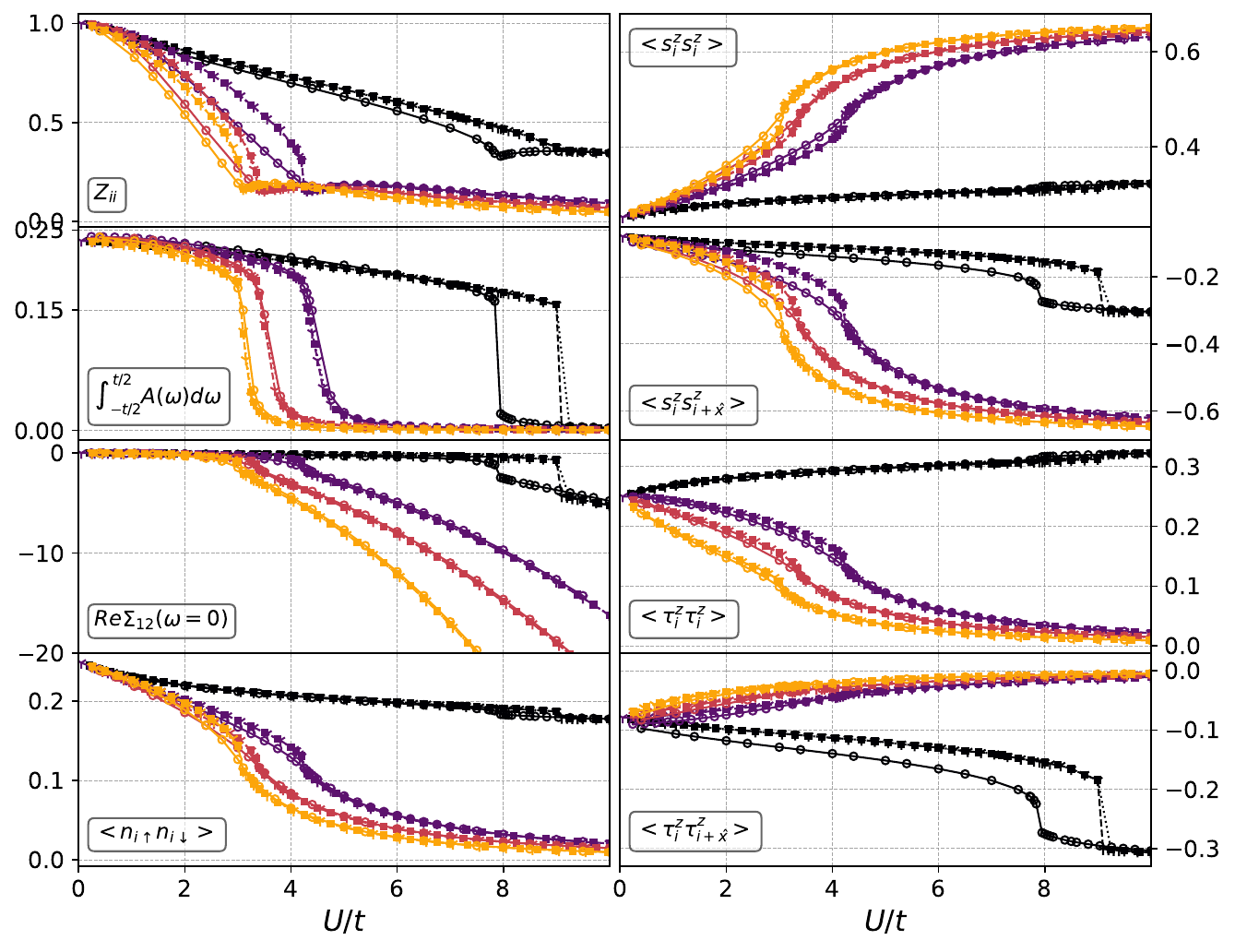}}
    \end{subfigure}
    \caption{Cluster quantities of the $1\times 2$ CDMFT solutions of the half-filled 2-band Hubbard model as a function of the interaction strength for different values of Hund's coupling $J$ with $N_b = 8$. \textbf{Left}: Quasiparticle residue of the impurity, spectral weight at low frequency in the local spectral function, real part of the off-diagonal self energy at zero frequency and double occupancy. \textbf{Right}: Local and nearest neighbor spin and band correlations.}
    \label{fig:1x2x2 J cluster observables}
\end{figure*}
We finally move to a two-band (two-orbital) Hubbard-Kanamori model model to analyze the role of a multi-orbital electronic structure. For this system we consider the two-site cluster to limit the growth of the Hilbert space. 

We will consider the case where the two bands have the same bandwidth  at half-filling. While this system has been studied extensively in single-site DMFT or using slave-particle methods ~\cite{koga_stability_2002,ono_mott_2003,lechermann_rotationally_2007,peters_orbital_2010,de_medici_hunds_2011,medici_modeling_2017}, the effect of non-local correlations the 2D case has received little attention, in particular very few studies have been done using cluster extensions of DMFT ~\cite{kita_spatial_2009,nomura_multiorbital_2014}, and to the best of our knowledge never at zero temperature. 

Again, for these calculations, we used ED in the case of $N_b=8$, and ASCI for $N_b=16$, with a typical number of $\sim 5\times 10^5$ determinants.
In Fig. \ref{fig:1x2x2 J cluster observables}, we first show the cluster quantities for this model for different values of the Hund's coupling $J$, with a the bath fixed to its smallest size $N_b=8$. 
Increasing the value of $J$ increases the degree of correlation, as we can see from the faster decrease spectral weight at low frequency and the quasiparticle residue. The observed trends with respect to the value of $J$ follow the trends identified in a variety of previous studies using  local approximations ranging from DMFT to auxiliary particles~\cite{koga_stability_2002,ono_mott_2003,lechermann_rotationally_2007,peters_orbital_2010,de_medici_hunds_2011,medici_modeling_2017,facio_nature_2017}, and variational Monte Carlo~\cite{de_franco_metal-insulator_2018}. In particular the critical $U$ for the Mott transition follows the predictions based on the atomic limit in which the gap is $ U-J$~\cite{de_medici_hunds_2011}.

The Mott transition takes place in a similar way to what we found for the 2-site cluster (see Fig. \ref{fig:1x2x1 cluster observables}). The suppression of spectral weight at low frequency is accompanied by the onset of a non-zero off-diagonal part of the self-energy, and an abrupt change of behaviour in the quasiparticle residue. Previous studies~\cite{lechermann_rotationally_2007,medici_modeling_2017,facio_nature_2017,de_franco_metal-insulator_2018} found the transition to be continuous at $J=0$, then first order for small finite values of $J/U$, and continuous again for values of $J/U \gtrsim 0.3$. Our results seem to meet the two latter expectations in the Irrep and Replica bath parameterizations, which are expected to be more reliable for a small number of baths. However,  at $J=0$, there appears to be a jump in our calculations for all computed observables except the quasiparticle residue which is consistent across the different bath parameterizations. 
Notice that, in the case  $J=0$, spin and orbital degrees of freedom have the same behavior, reflecting the $SU(4)$ symmetry of the system in this parameter regime (see Fig. \ref{fig:1x2x2 J cluster observables}).
    
As the value of J becomes finite, the symmetry of the system changes to $SU(2)\times SU(2)$, and the correlation functions display different behaviour: At strong values of the interaction, the double occupancies go to zero, the local spin moment is enhanced, and the band moment $\langle\tau^z_i\tau^z_i \rangle$ is reduced. Corresponding to an insulating system where each of the two bands is occupied by one electron, and the spins of these two electrons are aligned, as would dictate the atomic limit of the Kanamori Hamiltonian. Moreover, the nearest-neighbour correlation functions indicate increasing AFM ordering, and decreasing antiferro-orbital (AFO) ordering with increasing interaction. Therefore, the spin and band polarization are alternating in space, but this tendency is reduced in the bands as their occupation equilibrates and increased in the spin as each site becomes increasingly spin polarized.

\begin{figure*}
    \centering
    \begin{subfigure}{\textwidth}
    \subfloat{\includegraphics[width=0.9\linewidth]{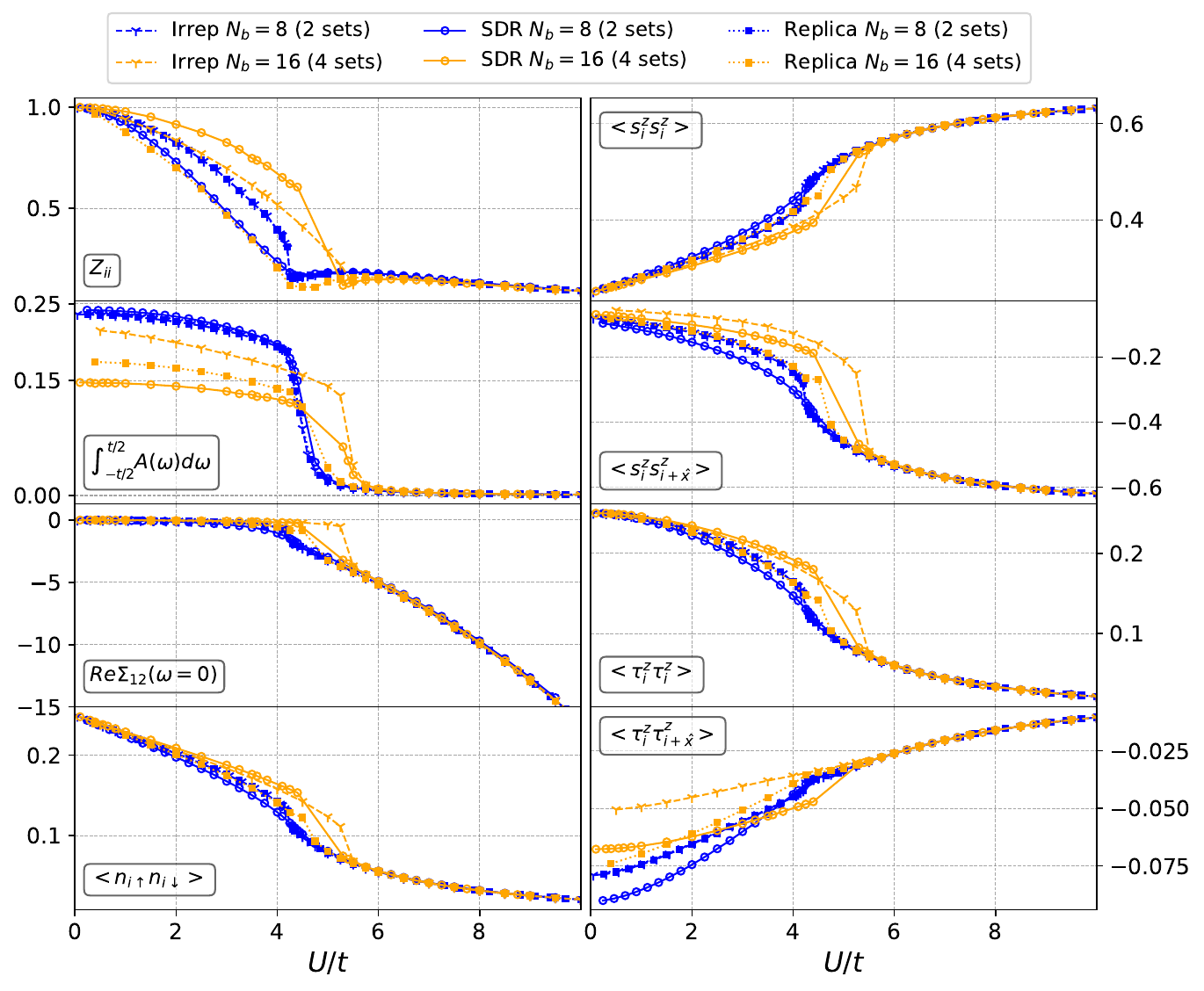}}
    \end{subfigure}
    \caption{Cluster quantities of the $1\times 2$ CDMFT solutions of the half-filled 2-band Hubbard model as a function of the interaction strength with $J=0.2U$. \textbf{Left}: Quasiparticle residue of the impurity, spectral weight at low frequency in the local spectral function, real part of the off-diagonal self energy at zero frequency and double occupancy. \textbf{Right}: Local and nearest neighbor spin and band correlations.}
    \label{fig:1x2x2 Nb cluster observables}
\end{figure*}

Here, we only tested the influence of an increasing number of bath sites for $J/U=0.2$, as shown in Fig.~\ref{fig:1x2x2 Nb cluster observables}. In this case, the results are very similar to what we found with the smallest number of bath sites, the main difference being an accentuation of the discontinuity at the transition, which was accompanied with an increasing difficulty to converge the calculations around the critical value for U. Nevertheless, unlike the single band Hubbard model calculations, in this system  a bath consisting $N_{\text{sets}}=4$ is not enough to reach full convergence with respect to the different bath parameterization for all the computed quantities at low to intermediate values of the interaction strength. It is for example difficult to pinpoint an unique value for the exact critical interaction value of the metal-insulator transition. This finding emphasizes the exceptional challenge of embedding in the multi-site, multi-orbital regime, and calls for the use of sophisticated solvers and bath parameterizations.
Nevertheless, all the calculations coincide for strong values of the interaction strength, when the system is fully insulating.

\section{Summary and Conclusion}

In this work, we have investigated the role of bath discretization—including bath size and parameterization in Hamiltonian-based embedding methods with multi-site (cluster) and  multi-orbital modelig. We considered in particular cluster dynamical mean field theory (CDMFT) solutions for two-dimensional single-band and two-band Hubbard models focusing on paramagnetic solutions at zero temperature. 
The use of the Adaptive Sampling Configuration Interaction (ASCI) solver enabled us to access bath sizes significantly larger than those typically achievable with Lanczos-based Exact Diagonalization (ED), thus allowing us to provide information about the convergence of the results with respect to size of the bath.

We compared three distinct bath parameterization schemes: two symmetry-inspired approaches (Irreps and Symmetry-Preserving Replica) and one parameterization designed to facilitate semidefinite relaxation (SDR) techniques in the fitting procedure. In all cases the bath is organized in sets containing the same number of sites of the impurity cluster.

For the single-band Hubbard model, we performed calculations using two-site and four-site cluster impurity models. Our results revealed that, at $N_{\text{sets}}=2$ which is the typical bath size in ED studies for the Hubbard plaquette, the results remained sensitive to the choice of bath parameterization. 
On the other hand, increasing the number of sets to four provides similar results for the various parameterizations, suggesting that such affordable baths are sufficient to limit the effects of the discretization.

Incidentally, our results display a first-order zero-temperature Mott-Hubbard transition for the single-band Hubbard model at half-filling for all parameterizations, in contrast with previous claims and in agreement with the  Variational Cluster Approximation (VCA). We believe however that a more systematic exploration of finite-bath effects is required to settle this question.

On the other hand, in the two-band Hubbard model, we performed a two-site cluster calculation. This demanding set-up has been used only in a few calculations, none of which -to the best of our knowledge- using zero temperature impurity solvers and with a Hamiltonian formulation. In this case, we still find a dependence on the bath parameterization even at the largest bath sizes that we tested (4 sets), probably because of the combined effect of the small size of the cluster and of the bath. 

Our calculations do not therefore allow to establish the order and the nature of the Mott transition in a conclusive and unbiased way. Nevertheless, qualitative observations can be made when using flexible solvers and bath parameterizations. The common trends in our calculations suggest, for example, that the order of the Mott transition may change with respect to single-site DMFT and other local methods as a result of the inclusion of non-local dynamical correlations.

These findings highlight the importance of both bath size and parameterization in Hamiltonian-based CDMFT, particularly when studying complex multi-orbital systems, which are necessary for material driven studies. However we find that for a plaquette one obtains converged results already for a bath composed by four sets of sites, while two-site clusters require larger baths.
Our calculations  also demonstrate that CDMFT combined with ASCI offers a powerful and flexible framework for exploring strongly correlated models beyond the reach of conventional ED approaches in this regime.

\begin{acknowledgments}
We ackwnoledge insightful discussions with G. Bellomia, S. Giuli and E. Linnèr. D.F.A. is also grateful to B. Bacq-Labreuil for useful discussions concerning the Irrep bath parameterization and for sharing some examples of its implementation in the Pyqcm library ~\cite{10.21468/SciPostPhysCodeb.23}.
We acknowledge financial support of MUR via PRIN 2020 (Prot. 2020JLZ52N 002) program, and  PRIN 2022 (Prot. 20228YCYY7), MUR PNRR Projects No. PE0000023-NQSTI, and No. CN00000013-ICSC.
\end{acknowledgments}

\newpage
\section{Appendix}
\appendix

\section{Symmetry Preserving Replica structure in the Irrep basis}
\label{Replica_in_Irrep_basis}

For the replica fit, we use the physical intuition that the DMFT bath is meant to approximate the effect of the rest of system on the impurity. Therefore, the bath parameters will have an internal structure that mimics the expected spatial symmetry of the solutions. For example, in the case of a single-band $2\times2$ impurity with 8 baths, we will have two sets (copies of the impurity) ($N_\text{sets}=2$) with internal onsite energies $\epsilon_\lambda$, nearest-neighbour hopping $t_\lambda$ and next-nearest-neighbour hopping $t'_\lambda$. This means that the matrix $\overleftrightarrow{{\epsilon}_\lambda}$ will not be diagonal, while the full matrix $\overleftrightarrow{{\epsilon}}$ will be a block diagonal matrix containing the different $\overleftrightarrow{{\epsilon}_\lambda}$.  

In addition, we must now define the coupling matrix, which will now have to satisfy the symmetries of the replica and the impurity. In our example, we will have the diagonal coupling $V_{0\lambda}$, the nearest neighbour coupling $V_{1\lambda}$, and the next nearest neighbour coupling $V_{2\lambda}$. In this case, the full matrix $\overleftrightarrow{{V}}$ will be a row of block matrices $\overleftrightarrow{V}_\lambda$  

\begin{figure}
    \centering
    \includegraphics[width=0.4\linewidth]{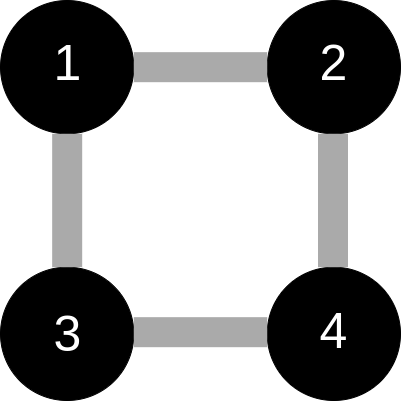}
    \caption{ Convention for the site indexing in the $2\times 2$ cluster .}
    \label{fig:2x2_convention}
\end{figure}

Let us explicitly build the matrices for this case using the convention shown in Fig. \ref{fig:2x2_convention} for the labeling of the sites in the cluster:
\begin{equation}
\overleftrightarrow{{\epsilon}_\lambda}=
    {\scriptstyle \begin{bmatrix*}[c]
 \epsilon_\lambda  & t_\lambda & t_\lambda &{t'}_\lambda \\
 t_\lambda & \epsilon_\lambda  & {t'}_\lambda & t_\lambda \\
 t_\lambda & {t'}_\lambda & \epsilon_\lambda  & t_\lambda \\
 {t'}_\lambda & t_\lambda & t_\lambda & \epsilon_\lambda  \\
    \end{bmatrix*}}
\end{equation}
and
\begin{equation}
    \overleftrightarrow{{V}_\lambda}=
    {\scriptstyle \begin{bmatrix*}[c]
 \text{V}_{0\lambda} & \text{V}_{1\lambda} & \text{V}_{1\lambda} & \text{V}_{2\lambda} \\
 \text{V}_{1\lambda} & \text{V}_{0\lambda} & \text{V}_{2\lambda} & \text{V}_{1\lambda} \\
 \text{V}_{1\lambda} & \text{V}_{2\lambda} & \text{V}_{0\lambda} & \text{V}_{1\lambda} \\
 \text{V}_{2\lambda} & \text{V}_{1\lambda} & \text{V}_{1\lambda} & \text{V}_{0\lambda} \\
    \end{bmatrix*}}
\end{equation}

Now, if we diagonalize $\overleftrightarrow{\epsilon_\lambda}$, we find
\begin{equation}
    \overleftrightarrow{{\epsilon}_\lambda}=\overleftrightarrow{U}  \overleftrightarrow{\tilde{\epsilon}_\lambda} \overleftrightarrow{U}^\dagger
\end{equation}
with

\begin{equation}
    \overleftrightarrow{\tilde{\epsilon}_\lambda}=
    {\scriptstyle \begin{bmatrix*}[c]
 \epsilon_\lambda -t_\lambda' & 0 & 0 & 0 \\
 0 & \epsilon_\lambda -t'_\lambda & 0 & 0 \\
 0 & 0 & -2 t_\lambda+t'_\lambda+\epsilon_\lambda  & 0 \\
 0 & 0 & 0 & 2 t_\lambda+t'_\lambda+\epsilon_\lambda  \\
    \end{bmatrix*}}
\end{equation}
and
\begin{equation}
    U=\frac{1}{2}    {\scriptstyle \begin{bmatrix*}[r]
        - & - & + & + \\ 
        + & - & + & - \\ 
        + & - & - & + \\ 
        + & + & + & +
    \end{bmatrix*}}.
\end{equation}
Moreover, in this diagonal basis for the baths, the couplings become 
\begin{equation}
\begin{split}
    \overleftrightarrow{    \tilde{V}_\lambda}&=\overleftrightarrow{{V}_\lambda}\overleftrightarrow{U}^\dagger\\
    &=
    {\scriptstyle \begin{bmatrix*}[r]
 \tilde{V}_E\bm{v}_{E(2)} & \tilde{V}_E\bm{v}_{E(1)} & \tilde{V}_{B2}\bm{v}_{B2} & \tilde{V}_{A1}\bm{v}_{A1} \\
    \end{bmatrix*}}
\end{split}
\end{equation}
where, $\bm{v}_\lambda$ are the basis vectors associated to each Irrep and the corresponding coefficients are given by $\tilde{V}_{A1}=\frac{\text{V}_{0\lambda}+2V_{1\lambda}+V_{2\lambda}}{{2} }$, $\tilde{V}_{B2}=\frac{\text{V}_{0\lambda}-2V_{1\lambda}+V_{2\lambda}}{{2} }$ and $\tilde{V}_{E}=\frac{\text{V}_{2\lambda}-V_{0\lambda}}{\sqrt{2}}$.

 We find, thus, that the symmetry preserving Replica structure for the bath is equivalent to the Irrep structure with one additional constraint: $\tilde{V}_{E_1}=\tilde{V}_{E_2}$.

\section{Conditions for a nematic bath}
\label{Replica constraint in irrep basis}
We consider again the case of the $2\times 2$ impurity with a single band, but we start from the irrep bath parameterization. The symmetry group  of the cluster is $c_{4v}$, and the irreducible representations (irreps) of the system are $A_1$,$B_2$ and $E$. In the basis of these irreps, the hybridization function must be block diagonal. We can thus write, in the basis of these irreps, 
\begin{equation}
    \Delta(i\omega) = \sum_\lambda^{N_\text{sets}}\sum_n^{N_{C}}  \frac{\bm{V}_{n}\otimes\bm{V}_{n}^\dagger}{i\omega-\epsilon_n}=\sum_\lambda^{N_\text{sets}}\sum_n^{N_{C}}  \frac{\delta^n}{i\omega-\epsilon_n},
\end{equation}
where each $\bm{V}_n$ gives the hybridization to the impurity orbitals of the $n$th bath orbital within set $\lambda$, and corresponds to one of the irreducible representations. Since the Irrep $E$ is of dimension 2, here there will be 2 bath sites associated to this irrep, sharing the same on-site energy $\epsilon_E$.
The irrep basis for this type of impurity is given by
\begin{equation}
\begin{split}
    \bm{v}_{A_1}=v_{A_1}    &(+,+,+,+)\\
    \bm{v}_{B_2}=v_{B_2}    &(+,-,-,+)\\
    \bm{v}_{E^{(1)}}=v_{E^{(1)}}    &(-,+,-,+)\\
    \bm{v}_{E^{(2)}}=v_{E^{(2)}}    &(+,+,-,-),
\end{split}
\end{equation}
where the cluster sites are ordered from left to right and from top to bottom (as shown in Fig. \ref{fig:2x2_convention}).

Now, using the irrep basis of the plaquette, and imposing block-diagonal quantities. We can have only four types of contributions to the hybridization function:
\begin{equation}
\begin{aligned}
    \delta^{A_1} = |v_{A_1}|^2 
    {\scriptstyle \begin{bmatrix*}[r]
        + & + & + & + \\ 
        + & + & + & + \\ 
        + & + & + & + \\ 
        + & + & + & +
    \end{bmatrix*}}, &\quad    \delta^{E^{(1)}} = |v_{E^{(1)}}|^2 
    {\scriptstyle \begin{bmatrix*}[r]
        + & - & + & - \\ 
        - & + & - & + \\ 
        + & - & + & - \\ 
        - & + & - & +
    \end{bmatrix*}},
     \\\delta^{B_2} = |v_{B_2}|^2 
    {\scriptstyle \begin{bmatrix*}[r]
        + & - & - & + \\ 
        - & + & + & - \\ 
        - & + & + & - \\ 
        + & - & - & +
    \end{bmatrix*}},
 &\quad
    \delta^{E^{(2)}} = |v_{E^{(2)}}|^2 
    {\scriptstyle \begin{bmatrix*}[r]
        + & + & - & - \\ 
        + & + & - & - \\ 
        - & - & + & + \\ 
        - & - & + & +
    \end{bmatrix*}}.
\end{aligned}
\end{equation}

In the symmetry-preserving replica bath implementation, we have the added constraint $v_{E^{(1)}}=v_{E^{(2)}}$, which means that in this case we will only have contributions generated from the set of matrices $\left\{\delta^{A_1},\delta^{B_2},\delta^E\}\right)$, with $\delta^E=\delta^{E^{(1)}}+\delta^{E^{(2)}}$, such that
\begin{equation}
\begin{aligned}
    \delta^{E} = |v_{E}|^2 
    {\scriptstyle \begin{bmatrix*}[r]
        + & 0 & 0 & - \\ 
        0 & + & - & 0 \\ 
        0 & - & + & 0 \\ 
        - & 0 & 0 & +
    \end{bmatrix*}}
    \end{aligned}
\end{equation}
Now, the subspace of matrices spanned by this set will only contain matrices of the form 
\begin{equation}
    \Delta = 
    \begin{bmatrix*}
        \alpha & \gamma & \gamma & \beta \\
        \gamma & \alpha & \beta & \gamma \\
        \gamma & \beta & \alpha & \gamma \\
        \beta & \gamma & \gamma & \alpha
    \end{bmatrix*},
\end{equation}

and so, the behavior of the hybridization function along the $x$ and $y$ directions will always be the same. We can then conclude that, in the symmetry-preserving replica implementation the breaking of the rotational symmetry associated with the nematic solution is not possible.
\FloatBarrier
\bibliographystyle{unsrt}
\bibliography{references}

\end{document}